\documentclass[a4paper,fleqn]{cas-sc}
\usepackage[authoryear]{natbib}
\usepackage[linesnumbered,ruled,vlined]{algorithm2e}
\usepackage{algpseudocode}
\usepackage{bm}
\usepackage{placeins}
\usepackage{subcaption}
\usepackage{graphicx}
\RestyleAlgo{ruled}

\usepackage{amsmath}
\usepackage{amsthm}

\theoremstyle{definition} 
\newtheorem{definition}{Definition}
\newtheorem{remark}{Remark}

\def\tsc#1{\csdef{#1}{\textsc{\lowercase{#1}}\xspace}}
\tsc{WGM}
\tsc{QE}
\tsc{EP}
\tsc{PMS}
\tsc{BEC}
\tsc{DE}

\begin{document}
\let\WriteBookmarks\relax
\def\floatpagepagefraction{1}
\def\textpagefraction{.001}

\shorttitle{Virtual Traffic Police: Large Language Model Augmented Traffic Signal Control for Unforeseen Incidents}

\shortauthors{Wei, Wang, and Yang}

\title [mode = title]{Virtual Traffic Police: Large Language Model-Augmented Traffic Signal Control for Unforeseen Incidents}                      



%
\author[1]{Shiqi Wei}[orcid=0009-0003-9100-9114]
\ead{shiqiwei@u.nus.edu}
\author[1]{Qiqing Wang}[orcid=0009-0003-0053-5910]

\ead{shiqiwei@u.nus.edu}




\author[1]{Kaidi Yang}[orcid=0000-0001-5120-2866]
\ead{kaidi.yang@nus.edu.sg}


\cormark[1]

\affiliation[1]{organization={Department of Civil and Environmental Engineering, National University of Singapore},
    addressline={1 Engineering Drive 2}, 
    city={Singapore},
    postcode={117576}, 
    country={Singapore}}

\cortext[cor1]{Corresponding author}



\begin{abstract}
Adaptive traffic signal control (TSC) has demonstrated strong effectiveness in managing dynamic traffic flows. However, conventional methods often struggle when unforeseen traffic incidents occur (e.g., accidents and road maintenance), which typically require labor-intensive and inefficient manual interventions by traffic police officers. Large Language Models (LLMs) appear to be a promising solution thanks to their remarkable reasoning and generalization capabilities. Nevertheless, existing works often propose to replace existing TSC systems with LLM-based systems, which can be (i) unreliable due to the inherent hallucinations of LLMs and (ii) costly due to the need for system replacement. To address the issues of existing works, we propose a hierarchical framework that augments existing TSC systems with LLMs, whereby a virtual traffic police agent at the upper level dynamically fine-tunes selected parameters of signal controllers at the lower level in response to real-time traffic incidents. 
To enhance domain-specific reliability in response to unforeseen traffic incidents, we devise a self-refined traffic language retrieval system (TLRS), whereby retrieval-augmented generation is employed to draw knowledge from a tailored traffic language database that encompasses traffic conditions and controller operation principles. Moreover, we devise an LLM-based verifier to update the TLRS continuously over the reasoning process. Our results show that LLMs can serve as trustworthy virtual traffic police officers that can adapt conventional TSC methods to unforeseen traffic incidents with significantly improved operational efficiency and reliability. 
\end{abstract}




\maketitle

\section{Introduction}
Traffic signal control (TSC) is a critical component in the management of urban transportation systems, providing a cost-effective way to mitigate congestion and reduce energy consumption without requiring road infrastructure modifications~\citep{li2023survey, wang2024traffic, wu2025big}. Researchers have developed a variety of TSC schemes, including pre-timed~\citep{wang2024traffic}, actuated~\citep{xu2021traffic}, and adaptive methods~\citep{varaiya2013max, han2016robust, hao2018model, van2019linear, tan2024privacy, ma2021deep, mo2022cvlight, li2024cooperative}.
In particular, adaptive TSC approaches have advanced significantly over the past decades, with notable progress in max-pressure control methods~\citep{varaiya2013max, tsitsokas2023two, xu2024smoothing}, optimization-based methods~\citep{han2016robust, hao2018model, van2019linear, pham2023distributed, tan2024privacy}, and learning-based methods such as reinforcement learning~\citep{chu2019multi, ma2021deep, mo2022cvlight, li2024cooperative}. These methods exploit data from massive roadside and mobile sensors to infer traffic states (e.g., traffic flow, queue length, and travel times) and make real-time control decisions (e.g., green ratios, cycle lengths, and offsets). Such adaptive paradigms can reduce vehicle delay under diverse and fluctuating traffic conditions. 

\begin{figure}[hbt!]
    \centering
    \includegraphics[width=0.45\linewidth]{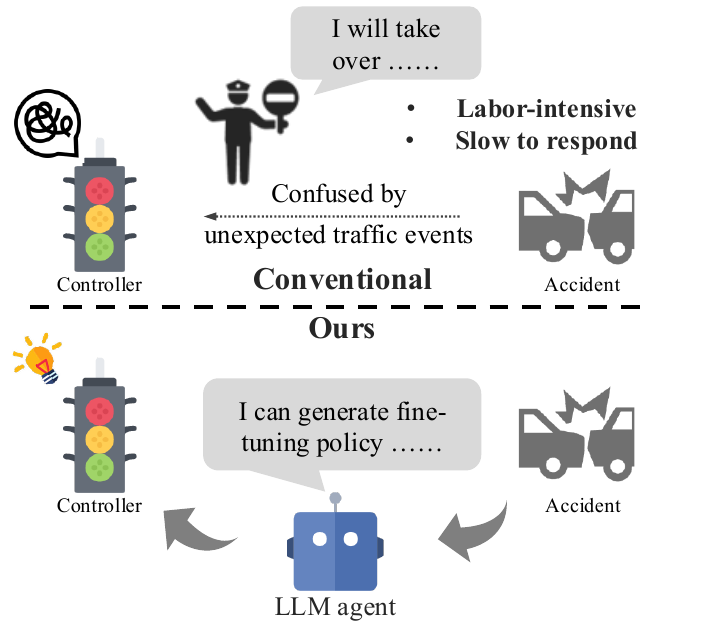}
    \caption{Comparison of conventional management of unforeseen traffic incidents and our proposed LLM-augmented framework}
    \label{fig:intro}
\end{figure}

However, existing adaptive TSC methods often struggle to perform efficiently when unforeseen traffic incidents occur, such as accidents and road maintenance. These methods are usually developed and evaluated under a single or limited set of incident types, and therefore fail to generalize to diverse unforeseen incidents encountered in real-world traffic environments \citep{chen2024emergency,mukhopadhyay2022spevms, tomforde2023incident,sheu2002stochastic}.
Current practice for handling traffic incidents typically requires manual interventions, whereby a traffic police officer is dispatched to the affected intersection to take over the current signal based on real-time conditions. Unfortunately, as shown in Figure \ref{fig:intro}, this process is both labor-intensive and slow to respond, often resulting in prolonged delays, safety risks, and increased operational costs for traffic management authorities. 

Recently, the rapidly developing foundation models, particularly large language models (LLMs), have shown remarkable capabilities in various tasks, ranging from natural language processing to complex problem-solving \citep{naveed2023comprehensive, liang2024survey}. LLMs have driven transformative progress across computer science, encompassing architectural innovation \citep{naveed2023comprehensive}, multi-modal reasoning \citep{cui2024survey}, robotics \citep{kim2024survey}, code generation \citep{jiang2024survey}, commonsense inference \citep{huang2022towards}, and information retrieval \citep{zhu2023large}. Their strong generalizability, rapid adaptability, and reasoning capability over structured and unstructured data present a timely opportunity for enhancing conventional intelligent transportation systems \citep{nie2025exploring, xu2024genai, long2024vlm}. Various approaches have utilized LLMs in applications such as autonomous driving \citep{mei2025llm, fang2025towards}, TSC \citep{wang2024llm, da2024prompt}, and spatio-temporal prediction \citep{he2025geolocation, nie2025joint}.

Although several attempts have explored the potential of LLMs in addressing traffic incidents in TSC \citep{wang2024llm, movahedi2024crossroads, lai2023llmlight, yuan2025collmlight,pang2024illm, da2024prompt}, existing works tend to use LLM agents to directly control traffic signals without considering their integration with conventional TSC methods. Nevertheless, directly leveraging LLMs for traffic signal control suffers from several limitations. 
First, entirely replacing existing TSC systems would require reinstalling or upgrading a large number of traffic control devices, leading to substantial installation costs.
Second, LLMs struggle to meet the stringent real-time requirements of high-frequency signal control. Their relatively long response times limit their effectiveness for direct intervention in phase transitions, which demand sub-second-level responsiveness. 
Third, LLMs are black-box models prone to hallucinations, potentially generating outputs that are factually incorrect or logically inconsistent with real-world traffic conditions~\citep{huang2025survey}.
Such hallucinations often originate in pre-training, arising from (1) the causal language modeling paradigm, where the model is trained to predict subsequent tokens solely based on preceding ones in a unidirectional manner, and (2) a limited knowledge boundary, which restricts the model’s ability to reason about rare, long-tail traffic events or copyright-sensitive information.
Notably, this limitation is unacceptable in a safety-critical and service-critical TSC system, as hallucinations could result in inappropriate signal phase allocations, such as assigning green time to blocked approaches, thereby even worsening traffic conditions, increasing spillback, and exacerbating traffic accidents.

Therefore, a reliable and cost-effective approach is to augment conventional model-based TSC systems with LLM agents (see Figure \ref{fig:intro}), which, to the best of our knowledge, has been rarely explored in the literature. 
It is worth noting that such integration is far from straightforward.
Simply coupling LLMs with existing controllers does not always guarantee reliability, especially when the system faces unforeseen traffic incidents that fall outside the distribution of training data or predefined rules. In such scenarios, LLMs may still produce unreliable outputs that compromise safety and efficiency.
To address this, a small portion of works~\citep{hussien2025rag,ding2024realgen,cai2024driving,yuan2024rag} have explored retrieval-augmented generation (RAG) to enhance the domain knowledge of LLM agents during inference. They incorporate domain-specific knowledge retrieved from databases, such as video embeddings paired with expert annotations in self-driving~\citep{yuan2024rag,dai2024vistarag} and contextual traffic-related common sense in TSC~\citep{zhang2024ragtraffic}.
Such efforts suggest that leveraging existing contextual traffic knowledge with the strong language understanding capabilities of LLMs can effectively enhance the reliability of LLM outputs in specific traffic-related tasks, offering valuable insights for advancing LLM-based TSC.
However, existing RAG-based methods still face two limitations. First, current approaches tend to build direct mappings from retrieved cases to high-level LLM outputs, without explicitly extracting intermediate traffic semantics such as traffic capacity reductions. This lack of structured semantic reasoning reduces the interpretability and generalizability of LLM outputs. Second, in practice, many traffic incidents are previously unseen, meaning that no sufficiently similar examples exist in the database to support reliable inference. 

To address the aforementioned challenges, we propose an LLM-augmented TSC framework that addresses two research questions (RQs):
\begin{description}
    \item[RQ1] \textit{How to augment conventional TSC systems with LLM agents?} 
To answer this RQ, we propose a hierarchical framework in which an LLM agent operates at the upper level to fine-tune key traffic parameters of the model-based TSC system at the lower level. 
This design combines the strong generalizability of LLMs over unstructured data with the strength of conventional TSC systems to enable high-frequency, real-time decision-making. 
\item[RQ2] \textit{How to enhance the reliability of LLM agents in response to unforeseen traffic incidents?} 
To answer this RQ, we design the LLM agent’s reasoning process to resemble that of an experienced traffic police officer with strong domain knowledge. Specifically, the agent is endowed with two key capabilities: (i) prior operational knowledge incorporated from historical incident reports, and (ii) the ability to continuously acquire and store new experience through real-world interactions.

\end{description}

\noindent \textbf{Statement of Contribution}. To summarize, our contributions are fourfold:
\begin{itemize}
    \item We propose a novel LLM-augmented TSC framework in which an LLM-based virtual traffic police agent enhances the reliability of conventional TSC methods in response to unforeseen traffic incidents. To the best of our knowledge, this is among the first efforts to integrate LLMs with conventional adaptive TSC methods.
    \item We devise a TRLS-based approach that enhances the LLM agent's domain knowledge by retrieving relevant contextual information from a tailored traffic language database constructed from historical traffic reports.
    \item We develop an LLM-based self-refinement mechanism that enables the LLM agent to act as both generator and verifier, allowing it to critique its own outputs and iteratively update the TLRS. This mechanism can continuously enhance the generalizability to previously unseen traffic scenarios.
    \item We evaluate the proposed framework using two adaptive TSC methods across four simulated cases, demonstrating its capability and effectiveness in improving traffic control performance under unforeseen traffic incidents.
\end{itemize}

The rest of this paper is organized as follows. 
Section~\ref{sec:Related work} reviews the related literature regarding incident-ware traffic control and LLM-enhanced traffic control.
Section~\ref{sec:Hierarchical Framework} presents our hierarchical methodological framework for the LLM-augmented TSC, which answers the first research question.
Section~\ref{sec:Self-Refined Traffic Language Retrieval System} introduces a self-refined TLRS for enhancing the reliability of the LLM agent, which answers the second research question.
Section~\ref{sec:case-study} performs case studies to evaluate our proposed method. Section~\ref{sec-con} concludes the paper.

\section{Related work}\label{sec:Related work}
In this section, we review the literature related to incident-aware traffic control in Section \ref{subsec: Incident-aware traffic control}. Then, we review the related work regarding LLM-based traffic control in Section \ref{subsec: Large language models for traffic control}. 

\subsection{Incident-aware traffic control}\label{subsec: Incident-aware traffic control}
Most existing work on traffic signal control (TSC) assumes that traffic flow follows a fixed or dynamic pattern within undersaturated or oversaturated ranges, without considering the impact of unforeseen traffic incidents. This limitation is shared by representative approaches, including actuated control, max-pressure, model predictive control (MPC), and reinforcement learning (RL). To address the limitations of conventional TSC in handling unforeseen events, recent studies have proposed incident-aware methods. Some of these studies focus exclusively on single-type incidents, such as emergency vehicle priority \citep{chen2024emergency,mukhopadhyay2022spevms} or lane-closure incidents \citep{tomforde2023incident,sheu2002stochastic}, which cannot be generalized to diverse traffic environments. Even though \citet{yue2023cooperative} considers various incident types, it primarily addresses post-incident traffic management, assuming that the occurrence and severity of incidents are exogenously specified.

\subsection{Large language models for traffic control}\label{subsec: Large language models for traffic control}
With the rise of LLMs, a growing body of research has explored their applications in traffic signal control (TSC). Recent studies \citep{wang2024llm,movahedi2024crossroads,lai2023llmlight,yuan2025collmlight} leverage the reasoning and generalization abilities of LLMs, positioning them as agents that directly control traffic signals. These works demonstrate that LLMs are capable of generating adaptive control policies in a zero-shot manner. However, approaches that rely solely on LLMs for direct control often overlook fundamental engineering challenges. First, the latency of LLM inference may hinder real-time responsiveness, which is a critical requirement in practical deployments. Second, pure LLM control may lead to unreliable decisions. For safety-critical intersections, such unsafe decisions could result in severe consequences, including traffic congestion or even accidents.

In addition, some works \citep{pang2024illm,da2024prompt} propose frameworks in which LLMs assess and refine reinforcement learning (RL) controllers. Nevertheless, by placing LLMs in the lower layer of the framework, these methods yield results that are not entirely trustworthy. Although LLMs are capable of interpreting decision objectives, they lack the capacity to concretely refine RL-based controllers. As a result, the strategies produced may still be unsafe, posing risks similar to those of purely LLM-based approaches.  

Meanwhile, in the autonomous driving domain, many studies \citep{long2024vlm,mei2025llm,fang2025towards,cai2024driving,yuan2024rag,luo2025senserag,cui2024board,xu2025tell,you2024v2x,chang2025driving,zhang2024wisead,liu2025language} have introduced LLMs for vehicle control. However, only a limited number of works \citep{long2024vlm,mei2025llm,fang2025towards} consider integrating LLMs with traditional controllers (e.g., model predictive control). To the best of our knowledge, no prior work in traffic control has investigated such a hierarchical design, which may offer a promising pathway to balance the reasoning capabilities of LLMs with the rigor and reliability of established control methods.

\section{A Hierarchical Framework for LLM-Augmented Traffic Signal Control}\label{sec:Hierarchical Framework}
This section introduces a general hierarchical framework for incident-aware LLM-augmented TSC, whereby LLMs complement, rather than replace, existing model-based solutions in transportation systems.
In Section \ref{subsec:problem}, we present the problem statement, highlighting the limitations of conventional TSC systems and motivating the potential role that LLMs can play in addressing these gaps.
In Section \ref{subsec:Overview}, we present a general overview of our proposed framework, which contains a virtual traffic police agent capable of leveraging the powerful reasoning and contextual understanding capabilities of LLMs.
It is important to note that this framework serves as a conceptual blueprint designed to address the limitations of conventional model-based control methods using LLMs. The proposed framework can be readily generalized to other decision-making tasks in transportation systems.

\subsection{Problem Statement}\label{subsec:problem}
Consider an isolated intersection with a set of incoming lanes $\mathcal{K}$. We choose isolated intersections as an initial building block that can be extended to large-scale urban transportation networks in future work. Let us denote the considered time horizon as a set of discrete intervals $\mathcal{T}=\{1,2,\cdots, T\}$ of a given size $\Delta t$. At each decision step $s \in \mathcal{T}$, an adaptive TSC algorithm $\mathcal{A}_{\bm{\theta}_s}$ parametrized by $\bm{\theta}_s$ is employed to make real-time signal timing decisions $\bm{u}_s$ based on observed traffic states (e.g., queue lengths) $\bm{x}_s = \{x_{s, k}\}_{k\in \mathcal{K}}$ of all incoming lanes, i.e., 
\begin{align}
    \bm{u}_s=\mathcal{A}_{\bm{\theta}_s}\left(\bm{x}_s\right)
\end{align}
where, in conventional TSC methods, the parameter $\bm{\theta}_s$ typically represents key traffic characteristics of the infrastructure, such as the saturation flow of each lane. Since the parameter is strongly correlated with the road topology and vehicle composition, it is usually predefined by traffic engineers based on historical experience. However, such static configurations often fail to adapt to unforeseen traffic incidents, leading to sub-optimal decisions.
For example, in the event of an accident causing a sudden drop in link capacity, an MPC-based approach assuming fixed link capacity may yield inaccurate system dynamics modeling and thus ineffective control strategies.

To address this limitation, our goal is to calculate a parameter $\bm{\theta}_s$ tailored  to the contextual description of the traffic incident $\bm{E}_s$ by minimizing the control objective $\mathcal{J}(\cdot)$, defined as 
\begin{align}
    \min_{\bm{\theta}_s \in \Theta} \mathcal{J}\left(\bm{x}_s, \mathcal{A}_{\bm{\theta}_s}(\bm{x}_s), \bm{E}_s\right) \label{eq:opt-conventional}
\end{align}
where the control objective $\mathcal{J}(\cdot)$ is a function with respect to the state $\bm{x}_s$, control decision $\bm{u}_s = \mathcal{A}_{\bm{\theta}_s}(\bm{x}_s)$, and contextual description $\bm{E}_s$ of the traffic incident. A formal definition of $\bm{E}_s$ is given in Definition \ref{dfn: incident description}. 
We make Remark \ref{rmk: Access to contextual descriptions} to discuss the accessibility of such contextual information.
The set $\Theta$ denotes the feasible set of the parameters $\theta_s$.

\begin{definition}[Contextual description $\bm{E}_s$ of the real-time traffic incident]\label{dfn: incident description}
At each decision step $s \in \mathcal{T}$, the contextual description $\bm{E}_s \in \mathcal{E}$ is defined as a structured tuple of semantic attributes characterizing the incident, which includes (i) the time information $\langle \text{Time} \rangle$, (ii) the location $\langle \text{Location} \rangle$ of the incident (i.e., at lane 1 from the west approach at intersection 1), and (iii) the type $\langle \text{Type} \rangle$ of the incident (i.e., car accidents or road maintenance).
\begin{align}
    \bm{E}_s = \langle \text{Time} \rangle \langle \text{Location} \rangle \langle \text{Type} \rangle
\end{align}
\end{definition}

\begin{remark}[Access to contextual description $\bm{E}_s$]
Our LLM-augmented TSC framework assumes that the contextual descriptions $\bm{E}_s$ of the real-time traffic incidents are available. This assumption is reasonable for two reasons. 
First, with the growing adoption of visual-language model-based solutions such as TrafficLens~\citep{arefeen2024trafficlens} and MAPLM~\citep{cao2024maplm}, such contextual information can be obtained from traffic cameras deployed across urban networks.
Second, even in cases where traffic cameras are unavailable at certain intersections, similar contextual descriptions can still be reported by drivers or passengers via vehicle-to-infrastructure (V2I) applications, a feature already implemented in some commercial mapping platforms, such as AMAP in China.
\label{rmk: Access to contextual descriptions}
\end{remark}

Conventionally, the optimization problem in Equation~\eqref{eq:opt-conventional} is formulated to address only a limited set of traffic incidents. In such cases, the contextual description $\bm{E}_s$ is restricted to a predefined set of incident types and directly encoded into the model constraints or objective function. This design, however, struggles to generalize to diverse and unforeseen incidents, as the space of possible incidents is far broader than what can be pre-specified. 

To overcome this limitation, we leverage the reasoning capability of LLMs to capture the complex relationship between $\bm{E}_s$ and the underlying traffic control parameters. Specifically, we aim to propose an LLM-powered agent that learns to identify a real-time fine-tuning policy $\mathcal{G}_{LLM}$, which dynamically determines $\bm{\theta}_s$ at each decision step $s$ in response to unforeseen traffic incidents:
\begin{align}
    \bm{\theta}_s \sim \mathcal{G}_{LLM}(\bm{E}_s).
\end{align}

In this formulation, we can leverage the strengths of LLMs to implicitly approximate the solution to the optimization problem~\eqref{eq:opt-conventional}, whereby the optimization problem is transformed into a token-based generative modeling problem. 
We use the operator $\text{LLM}(\cdot)$ to represent a pretrained LLM that maps prompts (e.g., event description) to designed output (e.g., parameter $\bm{\theta}_s$ and reasoning trajectories). 
Specifically, with input prompt $x$ that contains $\bm{E}_s$, we generate a token sequence of length $K$, i.e., $y=\left(y_1, y_2, \ldots, y_K\right)$, where $y_k \in \mathcal{V}$ and $\mathcal{V}$ denotes the token vocabulary that defines the complete set of discrete symbols (subwords, words, or characters). 
The LLM then models the conditional distribution over the output tokens $y$ as:
\begin{align}
P(y \mid x) = \prod_{k=1}^{K} P(y_k \mid y_{<k}, x; \psi),
\end{align}
where the model predicts each token $y_k$ step by step, conditioned on the context and the history of generated tokens $y_{<k}$. $\psi$ denotes the model parameters of the LLM. 

By reformulating the optimization task in this token-based manner, the LLM-powered agent can:  
(i) perform adaptive parameter updates guided by rich contextual descriptions $\bm{E}_s$, which are otherwise difficult to formalize into explicit constraints; and  
(ii) achieve generalizability to unforeseen incidents via zero-shot or few-shot reasoning, without the need for retraining on every new scenario. Nevertheless, token-based LLM generation also introduces inconsistency or hallucination to the generated output information, which may still lead to unreliable control parameter updates in our context. These issues will be addressed by our framework presented in Section~\ref{subsec:Overview}.

\subsection{Overview of the Hierarchical Framework }\label{subsec:Overview}
\begin{figure}[hbt!]
    \centering
    \includegraphics[width=0.95\linewidth]{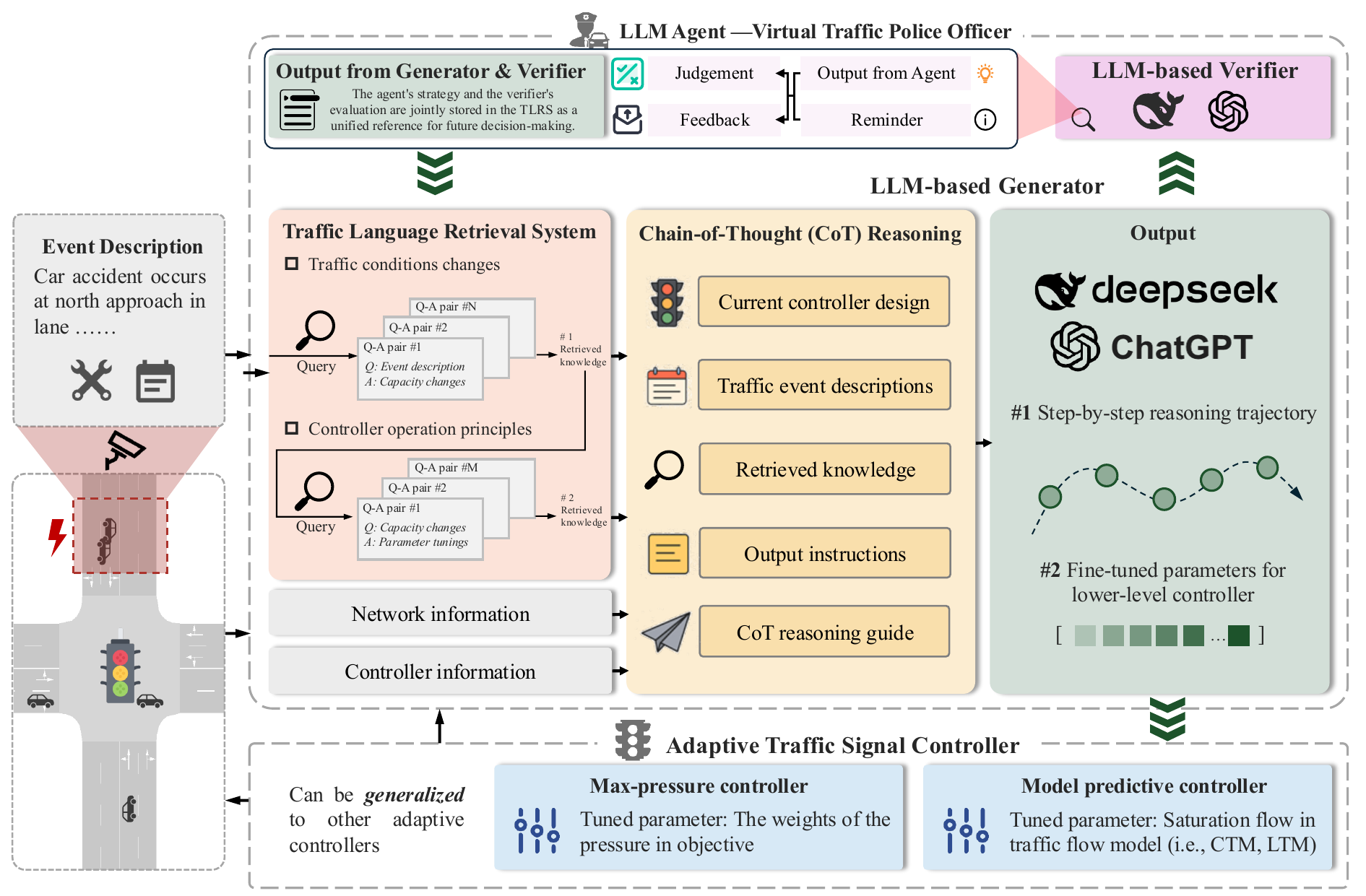}
    \caption{Illustration of our proposed incident-aware LLM-augmented TSC framework.}
    \label{fig:overview}
\end{figure}

The overview of our proposed incident-aware LLM-augmented TSC framework is illustrated in Figure \ref{fig:overview}. This hierarchical framework consists of two levels: (1) an upper-level LLM agent termed a \emph{virtual traffic police officer} that generates fine-tuned traffic parameters via an LLM-based policy $\mathcal{G}_{LLM}$ in response to unforeseen incidents (see Section \ref{subsubsec: Virtual Traffic Police Agent}), and (2) a lower-level adaptive traffic signal controller that executes the fine-tuned parameters in the adaptive traffic signal controller (see Section \ref{subsubsec:controller}). 
We next present the details of the two levels in our framework. It should be noted that this section focuses on the fundamental components of the hierarchical framework, which serve as the foundation for more advanced mechanism designs introduced in the next section.

\subsubsection{Virtual Traffic Police Agent}\label{subsubsec: Virtual Traffic Police Agent}
In this paper, we position the LLM agent at the upper level in the decision workflow, rather than allowing it to directly output control actions. In other words, the LLM agent serves as a high-level coordinator that interprets traffic incidents and infers their impacts on key traffic parameters.
Specifically, the policy of the agent $\mathcal{G}_{LLM}$ is composed of three key components as shown in Figure \ref{fig:overview}: (i) an LLM-based generator that produces reasoning trajectories and fine-tuned control parameters for the lower-level controller, (ii) a TLRS that retrieves relevant domain knowledge to ground the reasoning process and enhance decision reliability, and (iii) an LLM-based verifier to verify the generator’s outputs and iteratively update the TLRS, thereby enabling a self-refined and continuously improving decision-making framework. 
This section will focus on the first core component, the LLM-based generator, while the other two components will be introduced in Section \ref{subsec:TLRS} and Section \ref{subsec:Self-Refinement}, respectively.

In the LLM-based generator, we leverage the reasoning capability of LLMs by constructing zero-shot Chain-of-Thought (CoT) prompting \citep{wei2022chain, wang2023plan, kojima2022large}, which can help LLMs to perform zero-shot generalization to unseen scenarios. This kind of prompt engineering method is presented in Definition \ref{dfn:Zero-shot CoT prompting}.

\begin{definition}[Zero-shot CoT prompting \citep{kojima2022large}]\label{dfn:Zero-shot CoT prompting}
Zero-shot CoT prompting consists of two elements:  
(i) the task or problem description itself, and  
(ii) a simple trigger sentence such as \textit{"Let’s think step by step"}.  
The trigger sentence can also be extended into more detailed guidelines that explicitly specify the problem-solving procedure \citep{wang2023plan}.  
\end{definition}

At each decision step $s$, the LLM-based generator synthesizes four inputs as the CoT prompts: (i) a task overview $\bm{O}^g$ that includes intersection layout and signal controller information, (ii) the current traffic incident description $\bm{E}_s$, and (iii) the retrieved domain knowledge $\bm{R}_s$, and (iv) a structured instruction $\bm{I}^{g}$ that includes a reasoning step-by-step guidance and some reasoning reminders. 
The step-by-step procedure involves (i) understanding the traffic event and control objective, (ii) identifying the affected lanes, (iii) interpreting relevant domain knowledge from the historical event database, (iv) mapping the affected lanes to the parameter vector, and (v) finally determining the tuned parameters.
Then, an LLM processes all these inputs to generate the fine-tuned parameter $\bm{\theta}_s$ for the lower-level traffic signal controller, along with reasoning trajectories $\bm{\tau}_s = \{\bm{\tau}_s^{cond}, \bm{\tau}_s^{ctrl}\}$ that include the potential impact of the incident on traffic conditions, denoted as $\bm{\tau}_s^{cond}$, and the recommended controller operation, denoted as $\bm{\tau}_s^{ctrl}$,
\begin{align}
\label{eq:cot}
  \bm{\theta}_s,\bm{\tau}_s= \operatorname{LLM}(\bm{O}^g, \bm{E}_s, \bm{R}_s, \bm{I}^{g})
\end{align}

Then, the LLM processes all these inputs to generate the fine-tuned parameter $\bm{\theta}_s$, 
for the lower-level traffic signal controller, along with reasoning trajectories $\bm{\tau}_s$. 
Noted that the other two components of the LLM agent, the TLRS and LLM-based verifier, can be regarded as complementary modules that augment the input $\bm{R}_s$ of the generator by providing external knowledge grounding and iterative self-refinement, respectively. The extensibility of the agent design in response to unforeseen traffic incidents, enabled primarily by the generator, is further discussed in Remark~\ref{rmk: Extensibility of the Agent Design}.

\begin{remark}[Extensibility of the Agent Design]
It is important to note that the CoT prompts that contain four candidate inputs fed into the LLM-based generator can be flexibly extended to incorporate additional multimodal and contextually relevant information from transportation systems. For instance, the retrieved domain knowledge $\bm{R}_s$ can be enriched with structured traffic incident logs, simulation-based counterfactuals, and domain-specific ontologies to provide a more comprehensive knowledge base. Moreover, the task overview $\bm{O}^g$ can be enriched with visual data, such as intersection schematics or traffic camera snapshots. By enabling the integration of such heterogeneous sources, our CoT-based agent design facilitates richer contextual grounding and enhances the agent’s ability to reason about complex, real-world traffic scenarios.
\label{rmk: Extensibility of the Agent Design}
\end{remark}

\begin{figure}[!htbp]
    \centering
    \begin{subfigure}{0.48\textwidth}
        \centering
        \includegraphics[width=\linewidth,height=0.48\textheight,keepaspectratio]{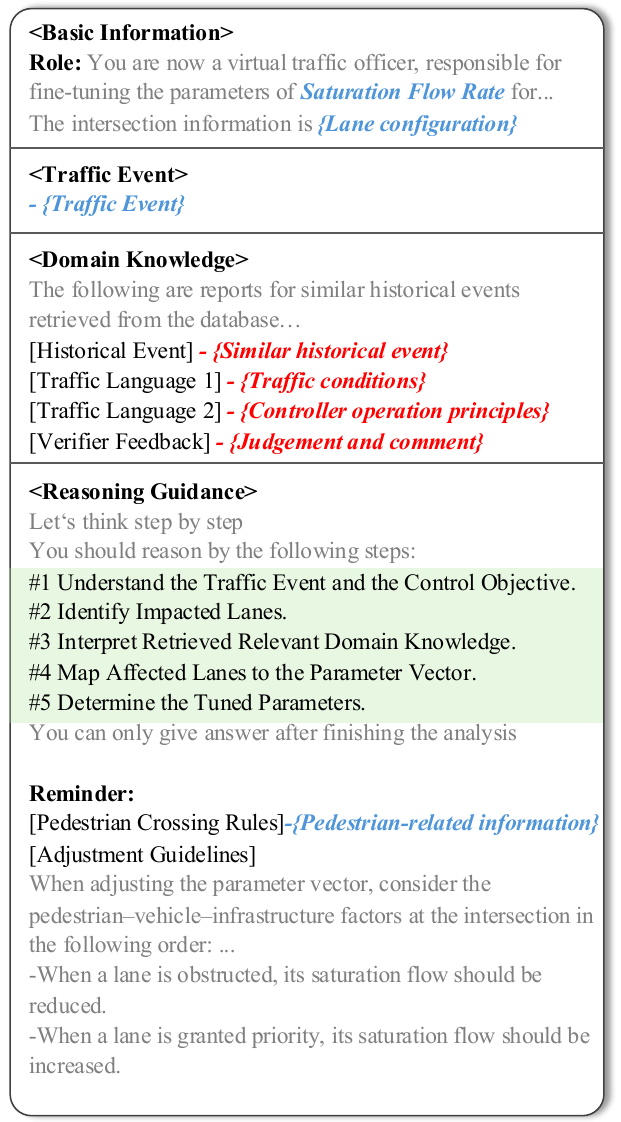}
        \caption{LLM-based Generator}
        \label{fig:CoT}
    \end{subfigure}%
    \hfill
    \begin{subfigure}{0.48\textwidth}
        \centering
        \includegraphics[width=\linewidth,height=0.48\textheight,keepaspectratio]{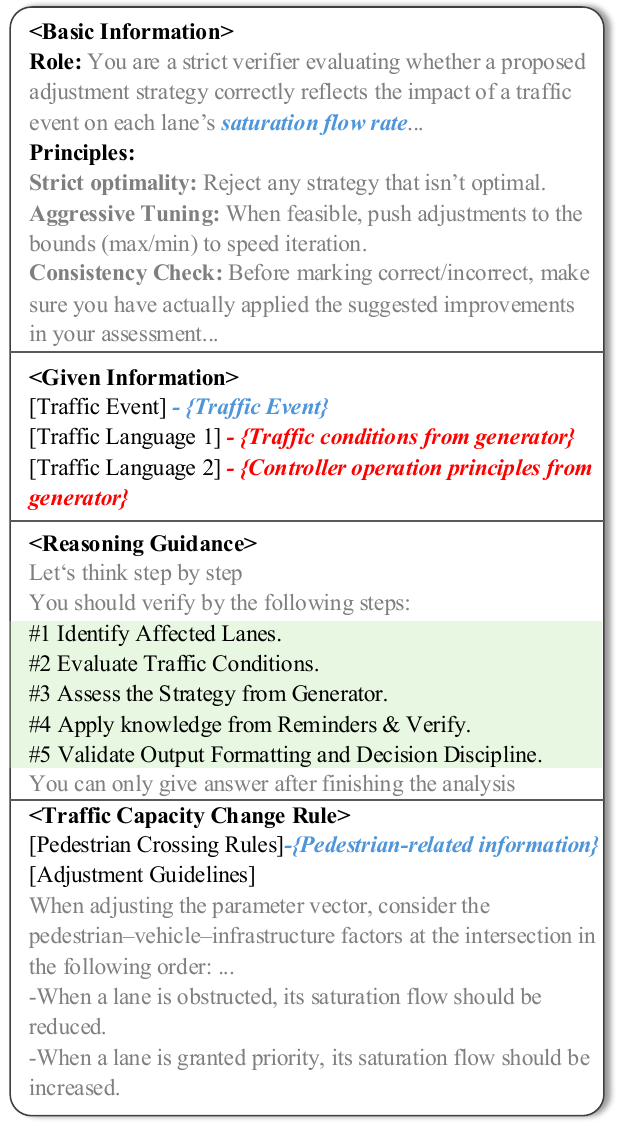}
        \caption{LLM-based Verifier}
        \label{fig:ver_CoT}
    \end{subfigure}
    
    \caption{Illustration of LLM-based generator and verifier prompts.}
    \label{fig:prompts}
\end{figure}

To this end, the structured prompt and reasoning guidance are shown in Figure \ref{fig:CoT}. Through a systematic step-by-step reasoning process with 5 steps, all information can be sufficiently processed to enable the LLM to identify the parameter $\bm{\theta}_s$, as well as the reasoning trajectories $\bm{\tau}_{cond}$ and $\bm{\tau}_{ctrl}$, that adapts the signal controller to real-time traffic incidents.
In Section \ref{sec:Self-Refined Traffic Language Retrieval System}, we will introduce how to prepare the retrieved domain knowledge $\bm{R}_s$ through a self-refined TLRS.

\subsubsection{Adaptive Traffic Signal Controller}\label{subsubsec:controller}
Once the upper-level LLM-based agent generates the fine-tuned parameter $\bm{\theta}_s$, the lower-level controller $\mathcal{A}_{\bm{\theta}_s}$ can adjust its key parameters to respond to unforeseen traffic incidents. This adaptive controller can be generalized to many conventional TSC methods with user-defined parameters. Here, we show how two widely adopted controllers (i.e., max-pressure control and model predictive control) can fit into this framework.

\noindent \textbf{Max-pressure controller:} At each decision step, the max-pressure controller for each intersection selects the next phase $p^{\star}$ based on
\begin{align} 
p^{\star} = \arg\max_{p\in\mathcal{P}} \sum_{(i,j)\in \mathcal{M}_{p}} c_{ij}w_{ij}
\end{align}
where $\mathcal{P}$ is a set of feasible signal phases, and $\mathcal{M}_{p}$ denotes the set of allowable movements during phase $p$, represented by incoming–outgoing lane pairs. The phase is selected based on the computed pressure $w_{ij}$ (e.g., the difference between downstream and upstream queue lengths) with pre-defined weights $c_{ij}$. 

The parameter $\bm{\theta}_s$ that needs to be tuned by the upper-level LLM agent includes the weights $\{c_{ij}\}_{(i,j)\in \mathcal{M}}$ with $ \mathcal{M} = \bigcup_{p \in \mathcal{P}} \mathcal{M}_p$. The weights $c_{ij}$ represent the \textit{saturation flow rates} of the corresponding lanes, reflecting their capacity under prevailing traffic conditions. For instance, when a lane is obstructed or constrained, the associated value of $c_{ij}$ decreases accordingly, indicating a reduction in saturation flow.

\noindent \textbf{Model predictive controller:} MPC-based controller relies on an embedded optimization problem, represented by Equations \eqref{eq:tsc_obj}-\eqref{eq:tsc_4}, that minimizes user-defined total cost within a finite time horizon in the future but only executes the action of the current time step.

\begin{align}
    \min_{\{\bm{u}_s\}_{s=t}^{t+H}}\quad& \sum_{s=t}^{t+H}F(\bm{x}_s,\bm{u}_s) \label{eq:tsc_obj}\\
    \mathrm{s.t.}\quad&  \bm{x}_{s+1} = f_{\bm{\phi}_s}(\bm{x}_s, \bm{u}_s),~\forall s = t, \cdots, t+H  \label{eq:tsc_1}\\
    & C\bm{u}_s =\bm{d},~\forall s = t, \cdots, t+H \label{eq:tsc_2}\\
    & u_{\min} \leq \bm{u}_s \leq u_{\max},~\forall s = t, \cdots, t+H \label{eq:tsc_3}\\
    & \bm{x}_t = \bm{\hat{x}}_t\label{eq:tsc_4}
\end{align}
where the objective function~\eqref{eq:tsc_obj} represents the control performance $F\left(\cdot\right)$ (e.g., total delay, fuel consumption, etc.) that depends on the traffic states $\bm{x}_t$ and signal timings $\bm{u}_t$ with a time horizon $H$.  
Constraint~\eqref{eq:tsc_1} represents the system dynamics $f_{\bm{\phi}_s}\left(\cdot\right)$, parametrized by $\bm{\phi}_s$, that describe the transitions between consecutive time steps. The detailed formats of $f_{\bm{\theta}_s}\left(\cdot\right)$ can be characterised by the cell transmission model (CTM), link transmission model (LTM), or store-and-forward model.  
Constraints~\eqref{eq:tsc_2}-\eqref{eq:tsc_3} impose restrictions on the signal timing parameters (e.g., green times) and constraint~\eqref{eq:tsc_4} sets the initial traffic states, where $\bm{\hat{x}}_t$ is the traffic states estimated by the observer.

The parameters $\bm{\theta}_s$ tuned by the upper-level LLM agent include the saturation flow vector $\bm{\phi}_s$ at time step $s$, similar to the definition in max-pressure control. More generally, $\theta_s$ may also represent traffic model parameters, such as the fundamental diagram in the cell transmission model (CTM) or the aggregate flow–density relation in macroscopic fundamental diagrams (MFDs), enabling flexible adaptation across different MPC formulations.

\section{Self-Refined Traffic Language Retrieval System}\label{sec:Self-Refined Traffic Language Retrieval System}
This section extends the hierarchical framework proposed in the previous section by devising two complementary mechanisms to enhance the reliability of the virtual traffic police agent at the upper level in response to unforeseen traffic incidents. First, we devise a TLRS in Section \ref{subsec:TLRS} to integrate domain knowledge into the agent's reasoning process during inference, thereby grounding the LLM’s outputs in historically validated incident-response patterns. Second, we develop an LLM-based verifier in Section \ref{subsec:Self-Refinement} to evaluate and refine the generated outputs from the LLM itself to update the TLRS continuously. This transforms the original TLRS into a self-refined retrieval system that empowers the LLM to adapt dynamically to diverse and safety-critical traffic scenarios.
These two mechanisms are specifically designed to enhance the quality and relevance of the retrieved knowledge $\bm{R}_s$ used in the LLM-based generator (see Section~\ref{subsubsec: Virtual Traffic Police Agent}), thereby improving the robustness and contextual reliability of the LLM-based agent under real-world operational conditions.

\subsection{Integrating Domain Knowledge via TLRS}\label{subsec:TLRS}
LLMs often struggle to answer domain-specific questions when directly applied to transportation systems \citep{xu2024genai}. To mitigate this limitation, we adopt the RAG technique \citep{fan2024survey} to build the TLRS, enabling the agent to retrieve domain knowledge and incorporate it into the LLM-based generator. Specifically, this system equips the LLM agent with expert references produced by professional traffic engineers, enhancing their understanding of (i) the impact of traffic incidents on traffic conditions, and (ii) the impact of traffic conditions on controller operation principles. To formalize such knowledge, we construct a traffic language database in Definition \ref{dfn: Traffic language database} based on expert experiences.

\vspace{-0.3em}
\begin{definition}[Traffic language database]\label{dfn: Traffic language database}
The traffic language database $\mathcal{D} = \{(d_1, J_1), (d_2, J_2), \ldots, (d_N, J_N)\}$ stores $N$ data entries, each including one chain of question–answer (Q-A) pairs $d_i$ and one evaluative signal $J_i$ (e.g., judgement and feedback on this chain). Specifically, each entry $d_i$ is represented as a two-stage Q-A chain: 
\begin{align}
    d_i =
        \bm{L}_i^{\text{inc}} \xrightarrow{\;\;\mathcal{O}_1\;\;} \bm{L}_i^{\text{cond}} 
        \xrightarrow{\;\;\mathcal{O}_2\;\;} \bm{L}_i^{\text{ctrl}},
\end{align}
where $\mathcal{O}_1$ and $\mathcal{O}_2$ denote the Q-A reasoning operators, i.e., $\bm{L}_i^{\text{cond}}=\mathcal{O}_1\big(\bm{L}_i^{inc}\big)$ and $\bm{L}_i^{\text{ctrl}}=\mathcal{O}_2\big(\bm{L}_i^{\text{cond}}\big)$.  
Specifically, $\mathcal{O}_1$ maps a traffic incident description $\bm{L}_i^{\text{inc}}$ to its impacts on traffic conditions $\bm{L}_i^{\text{cond}}$ (e.g., infrastructure capacity changes), while $\mathcal{O}_2$ translates the altered conditions $\bm{L}_i^{\text{cond}}$ into recommended controller operations $\bm{L}_i^{\text{ctrl}}$ (e.g., parameter tunings).  
\end{definition}
\vspace{-0.3em}

\begin{figure}
    \centering\includegraphics[width=0.75\linewidth]{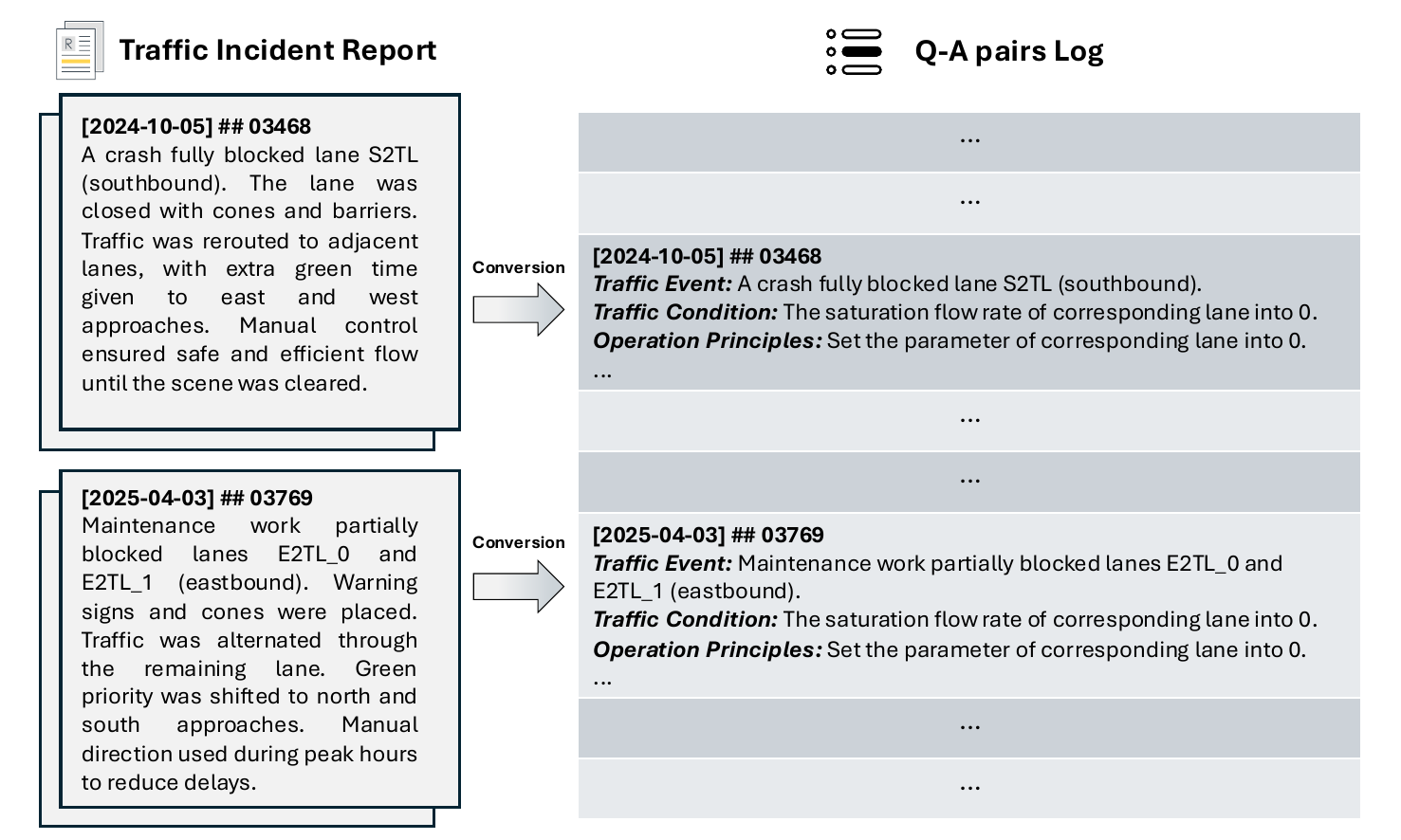}
    \caption{Examples of incident reports (left) and their corresponding Q-A pair representations (right) stored in the Traffic Language Retrieval System (TLRS).}
    \label{fig:report}
\end{figure}

We make the following remarks regarding the design considerations of the traffic language database. First, we incorporate traffic conditions into the traffic language database to formalize a chained Q–A pair. This is because traffic conditions serve as a universal traffic language across different types of traffic control scenarios. In addition, this design helps the LLM better interpret the contextual meaning of traffic incidents and translate them into appropriate control strategies.  
Second, we adopt the Q-A pairs design for the traffic language database, which can avoid tedious chunking processes in context retrieval \citep{gao2023retrieval} and enhance efficiency significantly. 
The transformation from a traffic incident report into a log of Q-A pairs can be automatically accomplished using LLMs, with illustrative examples provided in Figure~\ref{fig:report}.

Then, the retrieval system follows a two-step process to retrieve appropriate traffic language/knowledge from the references in the database. 
Let us define a unified traffic language retrieval operator $\operatorname{TrafficRetrieval}(\cdot)$,
\begin{align}
    \operatorname{TrafficRetrieval}(z ; m, \mathcal{D})
    = \left\{ (d_j, J_j) \in \mathcal{D} \;\middle|\;
        j \in \underset{i \in \{1,\cdots,N\}}{\arg \operatorname{top}_k}
        \big(\phi(z)^{\top} \phi(L_i^m)\big)
      \right\},\quad m \in \{\text{inc}, \text{cond}\}.
\end{align}
where the operator takes as input a query $z$, a retrieval stage signal $m$, a database $\mathcal{D}$ of chained Q–A pairs. 
Then, $\arg\operatorname{top}_k$ returns the top-$k$ entries in $\mathcal{D}$ whose associated traffic language $L_i^m$ is the most semantically similar to the query $z$, where we measure the cosine similarity between the questions and queries.
Function $\phi(\cdot)$ denotes the $\mathcal{L}_2$-normalized embedding function (e.g., nomic-embed-text \citep{nussbaum2024nomic}).

The proposed two-step retrieval process enables the retrieval of relevant information from the traffic language database, provided that the entries share semantically similar incident descriptions or comparable traffic conditions. Mathematically, with the operator $\operatorname{TrafficRetrieval}$ and the traffic incident description $\bm{E}_s$ at decision step $s$, the two-step retrieval becomes:
\begin{align}
    &\bm{R}_{s}^{\text{inc}} 
    = \operatorname{TrafficRetrieval}(\bm{E}_s; \text{inc}, \mathcal{D}), \label{eq:TrafficRetrieval-1}\\
    &\bm{R}_{s}^{\text{cond}} 
    = \bigcup_{(\bm{L}_i^{\text{inc}} \rightarrow \bm{L}_i^{\text{cond}} 
        \rightarrow \bm{L}_i^{\text{ctrl}}, J_i) \in \bm{R}_{s}^{\text{inc}}}\operatorname{TrafficRetrieval}(\bm{L}^{\text{cond}}_i; \text{cond}, \mathcal{D}), \label{eq:TrafficRetrieval-2}
\end{align}
where Equation \eqref{eq:TrafficRetrieval-1} defines the first-step retrieval, which retrieves a set $\bm{R}_s^{\text{inc}}$ of $k$ relevant items from $\mathcal{D}$ whose traffic incident descriptions are semantically similar to the query $\bm{E}_s$. 
Equation \eqref{eq:TrafficRetrieval-2} represents the second-step retrieval, which retrieves $\bm{R}_{s}^{\text{cond}}$ based on the first-step retrieval result $\bm{R}_s^{\text{inc}}$. Specifically, for each retrieved item $(d_i, J_i)$ in $\bm{R}_s^{\text{inc}}$ with $d_i=\bm{L}_i^{\text{inc}} \rightarrow \bm{L}_i^{\text{cond}} 
        \rightarrow \bm{L}_i^{\text{ctrl}}$, this step retrieves a set of $k$ relevant items from $\mathcal{D}$  whose traffic conditions are similar to the query $\bm{L}_i^{\text{cond}}$.        
        Note that the second-step retrieval augments the candidate set by including items associated with different incident descriptions but similar traffic conditions. This design arises from the observation that various incidents may produce similar traffic conditions and require analogous control responses.
        Therefore, we leverage the concatenation $\bm{R}_s=\left[\bm{R}_{s}^{\text{inc}}, \bm{R}_{s}^{\text{cond}}\right]$ to constitute the retrieved traffic language. The set $\bm{R}_s$ is then forwarded to the LLM-based generator to construct the CoT prompts, allowing the LLM to fine-tune controller parameters with a domain-aware prior.

\subsection{Self-Refinement via LLM-based Verifier}\label{subsec:Self-Refinement}
The TLRS introduced in the previous section equips the LLM agent with domain-specific exemplars and references that inform its reasoning, allowing the agent to reason similarly to an experienced traffic officer handling incidents based on prior experience.
However, the experiences stored in the TLRS are static and may not be updated in time to cover new or unseen traffic incidents. To address this limitation, we propose enhancing the TLRS with a self-refinement mechanism that continuously incorporates new experiences encountered during real-time agent–environment interactions. These new experiences will enable the agent to quickly adapt to unseen traffic scenarios. Specifically, the self-refinement mechanism extends the original traffic language database $\mathcal{D}$ after each decision step $s$ based on new chained Q-A pairs:
\begin{align}
    & d_s =  
        \bm{E}_{s} \rightarrow \bm{\tau}_{s}^{\text{cond}} \rightarrow 
        \bm{\tau}_{s}^{\text{ctrl}}, \\
    & \mathcal{D} \leftarrow \mathcal{D} \cup \{(d_s, J_s)\},
\end{align}
where $d_s$ denotes the new chained Q–A pair obtained from the LLM-based generator, which includes the incident description $\bm{E}_s$, the generated reasoning trajectory for traffic conditions $\bm{\tau}_{s}^{\text{cond}}$, and the reasoning trajectory for control principles $\bm{\tau}_{s}^{\text{ctrl}}$. 
The traffic language database $\mathcal{D}$ is then updated by incorporating the new entry $d_s$ together with its associated evaluative signal $\bm{J}_s$.

One key challenge in this refinement process lies in obtaining the evaluative signal $J_s$, which is typically unavailable due to the lack of human supervision or ground-truth annotations. To address this, we integrate an LLM-based verifier into the agent framework. This verifier assesses the generator’s outputs by tracing and evaluating their underlying step-by-step reasoning, enabling the system to perform self-verification without external annotation. Prior studies have been shown to be effective reasoners capable of self-verification \citep{weng2022large}. By leveraging this capability, the verifier ensures that only logically consistent experiences are incorporated into the traffic language database, thereby enabling continuous refinement of the TLRS over time. Specifically, after each decision step $s \in \mathcal{T}$, the LLM verifier evaluates the entire generation process and produces a contextual evaluative signal $J_s = \{\nu_{s}, \bm{a}_{s}\}$ that includes a binary judgment $\nu_{s}$ and feedback assessment $\bm{a}_{s}$ (i.e., reasons regarding the judgment):
\begin{align}
&\nu_{s}=\nu_{s}^{\text{cond}} \cdot \nu_{s}^{\text{ctrl}} \cdot \nu_{s}^{\text{map}} \label{eq:ver-1}\\
&\nu_{s}^{\text{cond}}, \nu_{s}^{\text{ctrl}}, \nu_{s}^{\text{map}}, \bm{a}_s= \operatorname{LLM}(\bm{O}^v, \bm{E}_s, \bm{\theta}_s, \bm{\tau}_s, \bm{I}^{v}) \label{eq:ver-2}
\end{align}
where Equation \eqref{eq:ver-1} decomposes the binary judgment $\nu_{s}$ into the product of several binary variables, including traffic condition verification $\nu_{s}^{\text{cond}}$, traffic control decision verification $\nu_{s}^{\text{ctrl}} $, and lane mapping verification $\nu_{s}^{\text{map}}$. Each variable takes the value $1$ if the verification passes and $0$ otherwise. Equation \eqref{eq:ver-2} obtains all verification outputs via an LLM-based verifier. Specifically, both fine-tuned parameters $\bm{\theta}_s$ and the reasoning trajectories $\bm{\tau}_s = \{\bm{\tau}_s^{\text{cond}}, \bm{\tau}_s^{\text{ctrl}}\}$ are included as the CoT prompts to be verified. In addition, the prompts also include a task overview $\bm{O}^v$, the current traffic incident description $\bm{E}_s$, and a structured instruction $\bm{I}^{v}$ that includes a reasoning step-by-step guidance and some reasoning reminders.

It is worth noting that the LLM-based verifier evaluates not only the correctness of the fine-tuned parameters but also the soundness of the whole reasoning trajectories, ensuring that both the decision logic and the resulting strategy satisfy the current traffic scenarios. This can be highlighted in the three decomposed binary variables in Equation \eqref{eq:ver-1}, which are detailed below to capture different aspects of the verification process.

\noindent \textit{1) Traffic condition verification}: 
The LLM-based verifier is instructed to first examine the generated traffic conditions $\bm{\tau}_s^{\text{cond}}$ in the reasoning trajectory $\bm{\tau}_s$. The traffic condition is a universal language representation in the transport world.
The verification is guided by domain knowledge \textit{<Traffic Capacity Change Rule>}, which specifies that the effective capacity of a lane should decrease when it is obstructed and increase when the lane is granted priority.
Accordingly, the binary judgment variable $\nu_s^{\text{cond}}$ is defined as
\begin{align}
    \nu_s^{\text{cond}} =
    \begin{cases}
    1, & \text{if the generated traffic condition aligns with \textit{<Traffic Capacity Change Rule>}}, \\
    0, & \text{otherwise}.
    \end{cases}
\end{align}

\noindent \textit{2) Traffic control decision verification}: 
Then, the LLM-based verifier evaluates whether the direction and magnitude of each element in the control vector $\bm{\theta}_s$ are appropriate. 
The strategy must reflect optimal control principles: aggressive (i.e., maximum or minimum) adjustments should be applied when justified, whereas moderate adjustments are expected under less severe conditions. 
The verifier explicitly checks for redundancy, omissions, overly conservative actions, or misalignment with the severity of the traffic event. 
Any deviation from these principles, such as suboptimal scaling or incorrect adjustment direction, is regarded as a violation of control knowledge. 
Formally,
\begin{align}
    \nu_s^{\text{ctrl}} =
    \begin{cases}
    1, & \text{if the generated traffic control decision conforms to the above principles}, \\
    0, & \text{otherwise}.
    \end{cases}
\end{align}

\noindent \textit{3) Lane mapping verification}: Finally, the LLM-based verifier verifies the index-to-lane mapping of each fine-tuned parameter, ensuring a precise alignment with the incident description. Only parameters corresponding to lanes affected by the traffic event are eligible for modification. The binary judgment for the lane mapping is represented as
\begin{align}
    \nu_s^{\text{map}} = \begin{cases}
    1, & \Theta_{\text{valid}} = \Theta_{\text{modified}}, \\
    0, & \text{otherwise}.
    \end{cases}
\end{align}
where $\Theta_{\text{valid}} = \{k \mid k \in \mathcal{K}_s^{\text{aff}}\}$, $\Theta_{\text{modified}} = \{k \mid \theta_s^k \neq \theta_{s-1}^k, k \in \mathcal{K}\}$, 
$\mathcal{K}_s^{\text{aff}}$ denotes the set of lanes affected by the current traffic incident, as inferred by the LLM agent.

\begin{figure}
    \centering
    \includegraphics[width=0.5\linewidth]{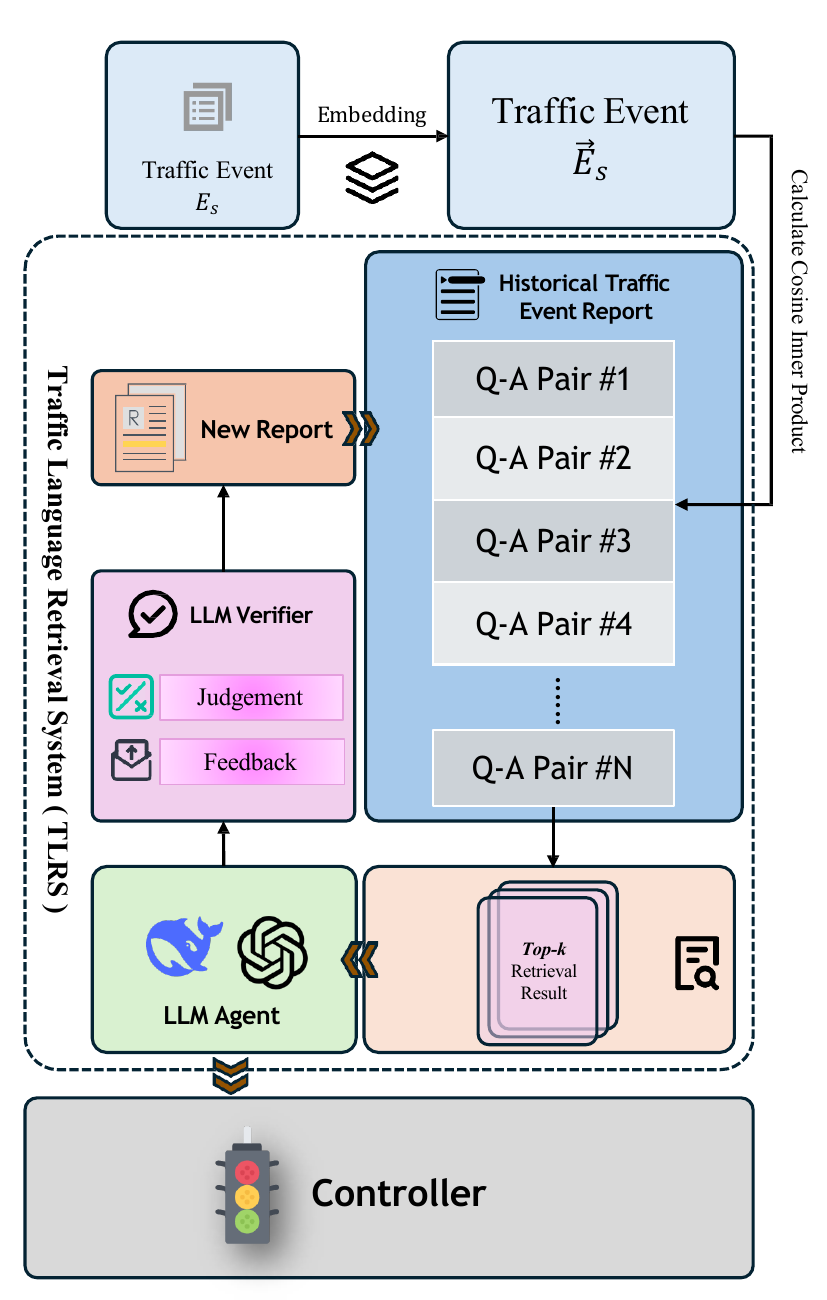}
    \caption{Overview of the Self-refinement Traffic Language Retrieval System (SRTLRS). The system retrieves top-$k$ relevant historical Q-A pairs based on the embedding of the current traffic event $E_s$. The LLM Agent generates a control strategy, which is then evaluated by the LLM Verifier through judgment and feedback. Verified outputs are used to assist the controller and are also added as new Q-A pairs to continually refine the retrieval system.}
    \label{fig:process}
\end{figure}

\begin{algorithm}[hbt!]
\caption{Incident-aware, LLM-augmented TSC with self-refined TLRS}
\label{alg:1}
\DontPrintSemicolon
\SetAlgoLined

\textbf{Input:} Traffic language database $\mathcal{D} = \{d_1, d_2, \ldots, d_N\}$, adaptive traffic signal controller's parameters $\bm{\theta}_1$, intervention frequency $Q$\\

\For{each time step $s = 1, 2, \ldots$}{
    \If{$s \bmod Q = 0$}{
        Obtain traffic incident description $\bm{E}_s$\;
        \textbf{TLRS:}\;
        Retrieve traffic domain knowledge $\bm{R}_s$ using Equations \eqref{eq:TrafficRetrieval-1}-\eqref{eq:TrafficRetrieval-2}\;
        \textbf{LLM-based generator:}\;
        $\bm{\theta}_s,\bm{\tau}_s= \operatorname{LLM}(\bm{O}^g, \bm{E}_s, \bm{R}_s, \bm{I}^{g})$\;
         \textbf{LLM-based verifier:}\;
        $\nu_{s}, \bm{a}_s= \operatorname{LLM}(\bm{O}^v, \bm{E}_s, \bm{\theta}_s, \bm{\tau}_s, \bm{I}^{v})$\;
        
        \textbf{Self-refined update:}\;
        $\bm{J}_s = \{\nu_{s}, \bm{a}_s\}$\;
        $\mathcal{D} \leftarrow \mathcal{D} \cup \{(d_s, \bm{J}_s)\}$\;
    }
    
    Observe traffic states $\bm{x}_s$\;
    \textbf{Adaptive traffic signal controller:}\;
    Make control decision $\bm{u}_s = \mathcal{A}_{\bm{\theta}_s}(\bm{x}_s)$\;
    
    Update $\bm{\theta}_{s+1} \leftarrow \bm{\theta}_s$\;
}
\end{algorithm}

\section{Numerical Experiment} \label{sec:case-study}
In this section, we perform case studies on four representative types of traffic incidents to evaluate the effectiveness of the proposed incident-aware LLM-agumented TSC framework. 
We first introduce experiment settings in Section \ref{subsec:Experiment settings}. Next, we assess the framework’s performance by comparing it with benchmarks under two scenarios: 1) when the TLRS contains relevant reference entries for the incident (Section~\ref{subsec:case-rel}), and 2) when no relevant references are available for retrieval (Section~\ref{subsec:case-norel}).

\subsection{Experiment settings}\label{subsec:Experiment settings}
This subsection introduces the experiment settings, including the traffic incidents setup, simulation configuration (i.e., intersection setup and traffic demand), and benchmark algorithms. All our experiments are simulated using a 60-minute  SUMO simulation \citep{lopez2018microscopic} during a typical morning peak.

\subsubsection{Traffic incidents setup}
\begin{figure}[htbp]
    \centering
    
    \begin{subfigure}{0.45\textwidth}
        \centering
        \includegraphics[width=\linewidth]{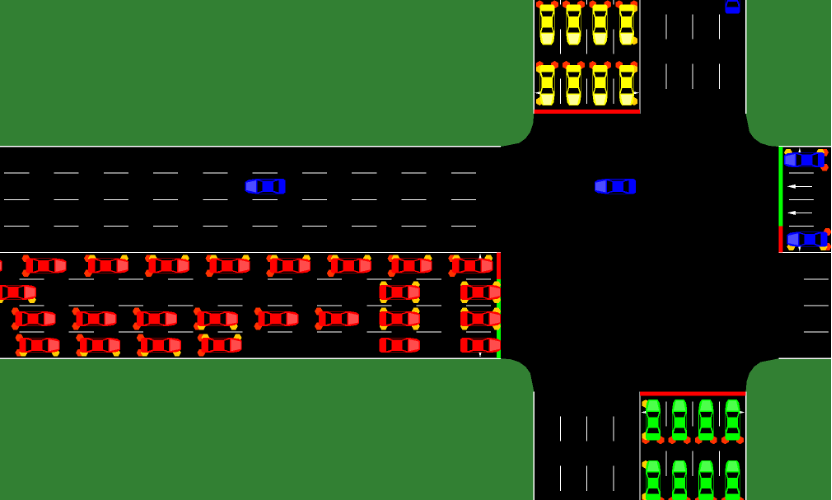}
        \caption{Accident}
        \label{fig:accident}
    \end{subfigure}
    \hfill
    \begin{subfigure}{0.45\textwidth}
        \centering
        \includegraphics[width=\linewidth]{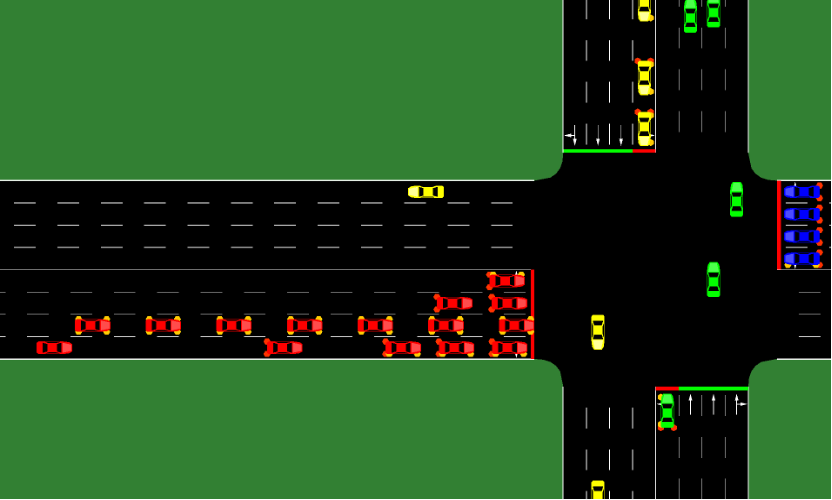}
        \caption{Road Maintenance}
        \label{fig:maintenance}
    \end{subfigure}
    
    \begin{subfigure}{0.45\textwidth}
        \centering
        \includegraphics[width=\linewidth]{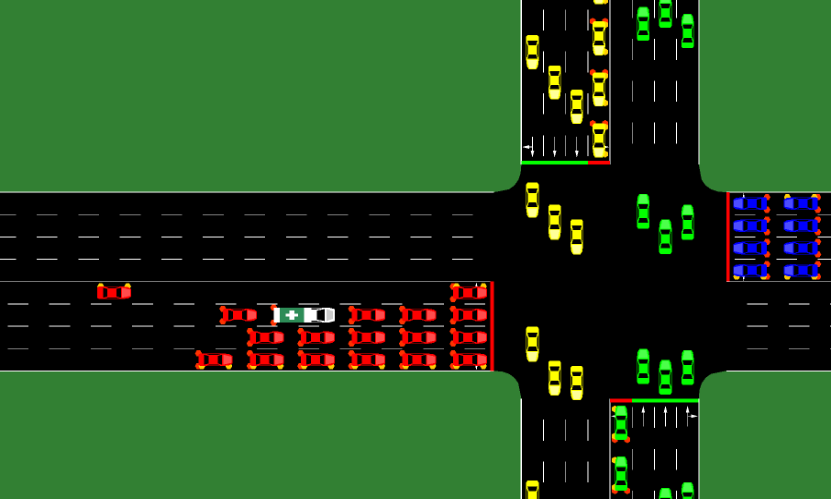}
        \caption{Ambulance Priority}
        \label{fig:ambulance}
    \end{subfigure}
    \hfill
    \begin{subfigure}{0.45\textwidth}
        \centering
        \includegraphics[width=\linewidth]{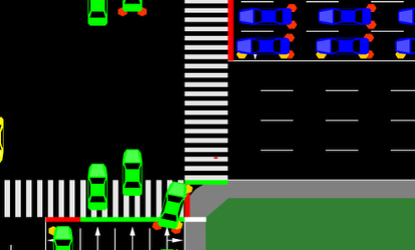}
        \caption{Elderly Crossing}
        \label{fig:elderly}
    \end{subfigure}
    
    \caption{Illustration of four incidents in SUMO}
    \label{fig:incidents_illu}
\end{figure}

In this paper, we consider four types of traffic incidents that may happen in an urban transportation network. Specifically, we consider two types of incidents related to lane blockage and two types of incidents related to priority passage. All of the traffic incidents may require manual intervention by the traffic police officers. The four representative traffic incident scenarios are illustrated in Figure \ref{fig:incidents_illu}.
\begin{itemize}
    \item \textbf{Car Accident (Complete Blockage):} A severe car accident occurs at a particular approach, resulting in a complete blockage of all lanes in that direction.
    \item \textbf{Road Maintenance (Partial Blockage):} Scheduled road maintenance takes place at a particular approach, causing partial lane closures and reduced capacity.
    \item \textbf{Emergency Vehicle Passage (Ambulance Priority):} The ambulance is passing through a particular approach, requiring temporary signal preemption or priority clearance (Ten ambulances are generated at regular intervals throughout the simulation).
    \item \textbf{Elderly Pedestrian Crossing:} An elderly pedestrian with a slow walking speed is about to cross, requiring an extended green phase to ensure safety.
\end{itemize}

We assume that only the \textbf{Car Accident} and \textbf{Road Maintenance} scenarios are common incident types at the studied intersection. As a result, the traffic language database contains relevant entries that can support the LLM’s reasoning process via the designed TLRS.
In contrast, for the \textbf{Emergency Vehicle Passage} and \textbf{Elderly Pedestrian Crossing} scenarios, no relevant historical records are available in the database. This setting is intended to simulate previously unseen incidents, thereby enabling us to evaluate the generalizability of the proposed LLM agent.

To evaluate the performance of the proposed framework across different types of incidents, we define the following metrics. For the \textbf{Car Accident} and \textbf{Road Maintenance} scenarios, the evaluation is based on the vehicle average delay (AD) and the average queue length (AQL) per lane.
In the \textbf{Emergency Vehicle Passage} scenario, we use the AD experienced by the emergency vehicle as the primary metric.
For the \textbf{Elderly Pedestrian Crossing} scenario, we perform multiple simulation runs and adopt the crossing completion rate (CCR) (i.e., the proportion of pedestrians successfully crossing during the green phase) as the evaluation metric. All of the above metrics can be obtained from the simulation outputs of SUMO.

\subsubsection{Simulation configuration}\label{subsubsec:sim-config}

\begin{figure}[!htbp]
    \centering
    \includegraphics[width=0.9\linewidth]{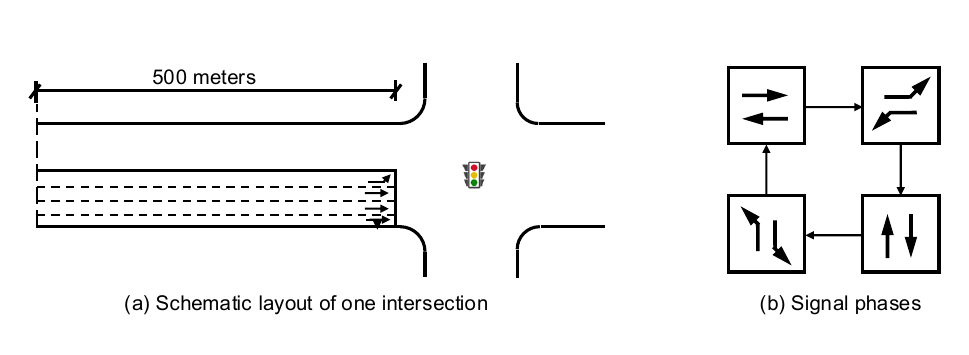}
    \caption{Illustration of simulated intersection.}
    \label{fig:sim}
\end{figure}

\noindent \textbf{Intersection setup}.  In this paper, we perform the numerical experiments on an isolated intersection, which consists of four bidirectional approaches, each comprising eight lanes.
The intersection operates under a four-phase scheme: (1) east-west through, (2) east-west left-turn, (3) north-south through, and (4) north-south left-turn, as shown in Figure \ref{fig:sim}. We employ two types of adaptive controllers: (1) Max-pressure and (2) MPC, both of which have been described in the Methodology (see Section \ref{subsubsec:controller}). The minimum green time is set to 15 seconds, the maximum green time to 45 seconds, and the yellow time to 3 seconds. Specifically, in the \textit{Elderly Crossing} scenario, pedestrian movements are considered. To ensure safe passage for pedestrians with a walking speed of 1.3 m/s, the minimum green time for both east-west and north-south through-right-turn phases is extended to 30 seconds \citep{mutcd_2009_4e06}.

\noindent\textbf{Traffic demand settings:}
To evaluate our methods under different traffic conditions, we consider two types of traffic demand: (1) an oversaturated scenario and (2) a moderate demand scenario. The oversaturated traffic volumes are presented in Table \ref{tab:oversaturated-demand}. The moderate demand scenario is configured as 60\% of the oversaturated volumes for each approach. To further quantify the demand intensity, we also compute the volume-to-saturation flow ($v/s$) ratios, where the saturation flow rate is assumed to be 1800 veh/h/lane. As shown in Table~\ref{tab:vs}, the total $v/s$ ratios are 0.93 and 0.56, which are consistent with our experimental settings for oversaturated and moderate demand levels, respectively.
For the \textbf{Elderly Pedestrian Crossing} scenario, we also introduced pedestrian flows into the simulation. The specific configuration is presented in Table \ref{tab:pedestrian-config}. 
We simulate the vehicle and pedestrian arrivals of each approach as a time-dependent Poisson process.

\begin{table}[!htbp]
\caption{Oversaturated Traffic Demand (vehicles/hour).}
\label{tab:oversaturated-demand}
\centering
\begin{tabular}{lcccc}
\toprule
\textbf{Approach} & \textbf{Left turn} & \textbf{Straight} & \textbf{Right turn} & \textbf{Total flow} \\
\midrule
East  & 413 & 900  & 187 & 1500 \\
West  & 495 & 1080 & 225 & 1800 \\
North & 298 & 1191 & 213 & 1701 \\
South & 251 & 1005 & 179 & 1435 \\
\bottomrule
\end{tabular}
\end{table}

\begin{table}[H]
\centering
\caption{Volume-to-saturation flow ($v/s$) ratios under different demand levels.}
\begin{tabular}{lcc}
\toprule
\textbf{Phase} & \textbf{Oversaturated} & \textbf{Moderate} \\
\midrule
EW Through  & 0.23 & 0.14 \\
EW Left     & 0.28 & 0.17 \\
NS Through  & 0.25 & 0.15 \\
NS Left     & 0.17 & 0.10 \\
\midrule
Total       & 0.93 & 0.56 \\
\bottomrule
\end{tabular}
\label{tab:vs}
\end{table}

\begin{table}[htbp]
\caption{Pedestrian Flow Settings.}
\label{tab:pedestrian-config}
\centering
\begin{tabular}{lcc}
\toprule
\textbf{Direction} & \textbf{Normal Pedestrian(peds/hr)}& \textbf{Elderly Pedestrian(peds/hr)}\\
\midrule
East–West   & 300 & 0 \\
West–East   & 300 & 0 \\
North–South & 300 & 15\\
South–North & 300 & 15\\
\midrule
\textbf{Walking Speed (m/s)} & 1.3 & 0.7 \\
\bottomrule
\end{tabular}
\end{table}

\subsubsection{Benchmark algorithms}\label{subsubsec:Benchmark algorithms}

We compare the performance of our framework with the following benchmarks:

\begin{itemize}
    \item \textbf{TSC:} This benchmark uses a conventional traffic signal controller (e.g., max-pressure controller and MPC) without the assistance of LLMs, which serves as the baseline.
    
    \item \textbf{LLM-augmented TSC (TSC-LLM):} This benchmark uses an LLM-based generator to generate fine-tuned parameters for the base controller without domain knowledge enhancement through the TLRS.
    \item \textbf{LLM-augmented TSC without CoT prompting (TSC-LLM without CoT):} This benchmark uses an LLM-based generator to generate fine-tuned parameters. However, the prompts only contain the basic task overview  $\bm{O}^g$ and the current traffic incident description $\bm{E}_s$.
    \item  \textbf{LLM-augmented TSC with self-refined TLRS (TSC-LLM-TLRS):} Our proposed framework that extends TSC-LLM by integrating a self-refined TLRS. This system enhances the LLM agent with domain-specific historical knowledge to guide its outputs. We incorporate $n=6$ historical references into the database, which simulates historical incident reports recorded by traffic experts/engineers.
    It should be noted that this traffic language database only contains relevant references regarding \textbf{car accident} (three references) and \textbf{road maintenance} (three references).
    Moreover, the traffic incidents recorded in the traffic language database differ from our test traffic incidents in geographic layout, number of affected lanes, and textual description. 
\end{itemize}

We conduct experiments using the API of two currently available, advanced, and representative LLMs:
\begin{itemize}
\item DeepSeek-V3 (671B): A powerful model known for its exceptional reasoning and contextual understanding capabilities.

\item ChatGPT-4o: A widely used multimodal LLM with strong performance across a broad range of tasks, capable of handling most real-world scenarios effectively.
\end{itemize}

For the LLM Agent, we set the temperature to $1$ to encourage diverse output generation. In contrast, the temperature for the LLM Verifier is set to $0$ to ensure stability and consistency in its evaluations.

\subsection{Performance comparison on incidents with relevant references}\label{subsec:case-rel}
This subsection discusses the model performance under the traffic incidents, \textbf{Car Accident} and \textbf{Road Maintenance}. For these two incidents, there already exists relevant references in the traffic language database in our proposed framework, TSC-LLM-TLRS. We analyze the performance of our proposed framework in two aspects. First, we verify the rationality of the LLM inference process under two benchmark settings: TSC-LLM and TSC-LLM-TLRS. Specifically, we verify the validity of the LLM agent’s retrieved traffic language and output in each benchmark. This step ensures that the LLM agent is capable of reasoning about the current traffic conditions in a manner similar to a human traffic police officer. Second, we assess the resulting traffic performance by comparing four benchmarks introduced earlier, using the corresponding evaluation metrics. This analysis aims to explore (i) how the LLM Agent with Chain-of-Thought (CoT) module outperforms the LLM Agent without CoT module in decision-making quality, and (ii) whether the external domain-specific knowledge from the TLRS can improve the overall framework performance.

\begin{figure}
    \centering
    \includegraphics[width=0.9\linewidth]{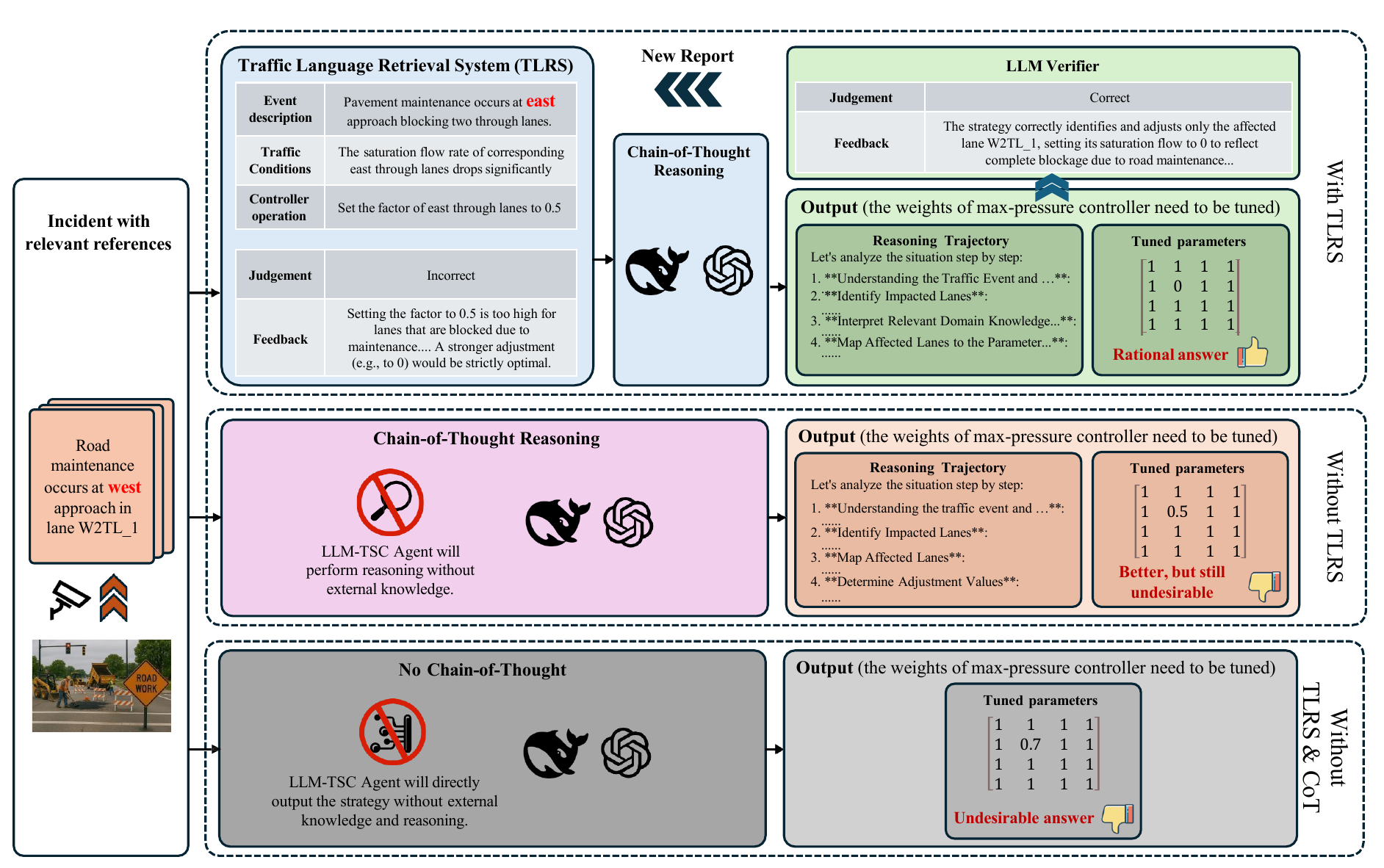}
    \caption{A case that demonstrates the inference process of the LLM Agent when handling an incident with relevant references.}
    \label{fig:case1}
\end{figure}

\noindent\textbf{Demonstration of LLM inference process:} Figure \ref{fig:case1} illustrates an example scenario where the incident description says \textit{"a road maintenance occurs at the west approach"}, resulting in the complete blockage of a lane. The max-pressure controller is selected as the lower-level controller, and we compare the LLM inference process among TSC-LLM without CoT, TSC-LLM, and TSC-LLM-TLRS. In our proposed framework \textbf{TSC-LLM-TLRS}, the TLRS successfully retrieves two relevant references, which include:(1) changes in traffic conditions (e.g., \textit{"the saturation flow rate of the corresponding lane drops to zero"}), and (2) control principles (e.g., \textit{"set the parameter of the corresponding lane to 0"}), as shown in blue. With this retrieved information, the LLM Agent is able to generate fine-tuned parameters and a detailed reasoning trajectory, as shown in green. It correctly identifies the blocked lanes and sets their pressure weights to zero, which aligns with proper control response under such an unforeseen incident.
It is important to note that although the referenced historical incidents are not identical to the current one (e.g., they occur on different approaches or the lanes that are affected are different), the framework still generates rational parameter adjustments for the controller.

In contrast, under the \textbf{TSC-LLM} benchmark, the agent fails to produce a reasonable adjustment strategy that the blocked lanes are still assigned non-zero weights (e.g., 0.5 in TSC-LLM and 0.7 in TSC-LLM without CoT), which may result in suboptimal control decisions at the lower level. This example clearly demonstrates that \textbf{TSC-LLM-TLRS} exhibits stronger generalization and reliability for unforeseen traffic incidents.

\noindent\textbf{Performance Comparison:} Table~\ref{tab:controller_comparison} presents the performance comparison between our proposed method and other benchmarks under two demand levels. The results demonstrate the effectiveness of our LLM-augmented framework in improving traffic control performance in response to traffic incidents, as evidenced by the significant reductions in AD and AQL across various scenarios.

\begin{table}[htbp]
\centering
\footnotesize 
\caption{Performance comparison under car accidents and road maintenance scenarios.}
\label{tab:controller_comparison}
\resizebox{\linewidth}{!}{
\begin{tabular}{@{}ccc >{\centering\arraybackslash}p{3.5cm} cc cc@{}}
\toprule
\textbf{Lower-level controller} & \textbf{Demand level} & \textbf{Methods} & \textbf{LLM} & \multicolumn{2}{c}{\textbf{Car accident}} & \multicolumn{2}{c}{\textbf{Road maintenance}} \\
& & & & AD (s) & AQL (m) & AD (s) & AQL (m) \\
\midrule

\multirow{12}{*}{\textbf{Max-pressure}}& \multirow{6}{*}{Oversaturated}& Baseline & - & 305.51 & 338.01 & 140.19 & 228.72 \\
& & \multirow{2}{*}{TSC-LLM without CoT}& Deepseek-V3 & 262.81 & 308.32 & 139.44 & 229.36 \\
& &                              & ChatGPT-4o & 262.81 & 308.32 & 139.44 & 229.36 \\
& & \multirow{2}{*}{TSC-LLM}& Deepseek-V3 & 262.81 & 308.32 & 139.17 & 227.57 \\
& &          & ChatGPT-4o & 262.81 & 308.32 & 139.44 & 229.36 \\

& & TSC-LLM-TLRS & Deepseek-V3/ChatGPT-4o & 232.33 & 288.69 & 134.67 & 221.14 \\

\cmidrule(lr){2-8}& \multirow{6}{*}{Moderate}& Baseline & - & 213.43 & 154.79 & 28.54 & 25.11 \\
& & \multirow{2}{*}{TSC-LLM without CoT}& Deepseek-V3 & 191.92 & 139.83 & 27.48 & 24.60 \\
& &                              & ChatGPT-4o & 191.92 & 139.83 & 27.48 & 24.60 \\
& & \multirow{2}{*}{TSC-LLM}& Deepseek-V3 & 191.92 & 139.83 & 27.82 & 24.83 \\
& &          & ChatGPT-4o & 191.92 & 139.83 & 27.48 & 24.60 \\
& & TSC-LLM-TLRS & Deepseek-V3/ChatGPT-4o & 177.88 & 133.41 & 25.90 & 23.60 \\

\midrule

\multirow{12}{*}{\textbf{MPC}}& \multirow{6}{*}{Oversaturated}& Baseline & - & 215.88 & 304.09 & 135.88 & 240.55 \\
& & \multirow{2}{*}{TSC-LLM without CoT}& Deepseek-V3 & 190.15 & 293.15 & 136.93 & 234.26 \\
& &                              & ChatGPT-4o & 190.15 & 293.15 & 124.75& 236.57\\
& & \multirow{2}{*}{TSC-LLM}& Deepseek-V3 & 190.15 & 293.15 & 126.57 & 214.04 \\
& &          & ChatGPT-4o & 190.15 & 293.15 & 136.93 & 234.26 \\

& & TSC-LLM-TLRS & Deepseek-V3/ChatGPT-4o & 185.28 & 274.25 & 121.62 & 218.44 \\
\cmidrule(lr){2-8}& \multirow{6}{*}{Moderate}& Baseline & - & 197.56 & 151.58 & 54.84 & 45.47 \\
& & \multirow{2}{*}{TSC-LLM without CoT}& Deepseek-V3 & 186.38 & 140.85 & 59.75 & 48.71 \\
  & &                              & ChatGPT-4o & 186.38 & 140.85 & 59.75 & 48.71 \\
& & \multirow{2}{*}{TSC-LLM}& Deepseek-V3 & 186.38 & 140.85 & 55.23 & 46.54 \\
& &          & ChatGPT-4o & 186.38 & 140.85 & 59.75 & 48.71 \\
& & TSC-LLM-TLRS & Deepseek-V3/ChatGPT-4o & 166.22 & 128.45 & 49.49 & 41.16 \\

\bottomrule
\end{tabular}
}

\end{table}

Under oversaturated demand, our method, TSC-LLM-TLRS, achieves the most significant improvements. In the car accident scenario, it can reduce the AD produced by the max-pressure controller from 305.51 seconds to 232.33 seconds, with more than 23\% improvement. When using the MPC controller, TSC-LLM with SRTLRS is still able to reduce AD from 215.88 seconds to 185.28 seconds. Similarly, in the road maintenance scenario, TSC-LLM-TLRS also outperforms other methods. Overall, the performance of TSC-LLM-TLRS consistently outperforms the other three methods under these two scenarios.

Meanwhile, the performance of our proposed method remains consistent across both Deepseek-v3 and GPT-4o, which can be attributed to the self-refined TLRS component that enriches the LLM with domain knowledge. This shows that LLM can reason based on retrieved similar historical incidents and generate empirically grounded strategies. These findings underscore the potential of integrating RAG techniques to enhance operational efficiency and decision reliability. In summary, these results show that the integration of traffic-language retrieval and LLM reasoning in our TSC framework can effectively fine-tune conventional TSC parameters in real time.

\begin{figure}[!htbp]
    \centering
    
    \begin{subfigure}{0.45\textwidth}
        \centering
        \includegraphics[width=\linewidth]{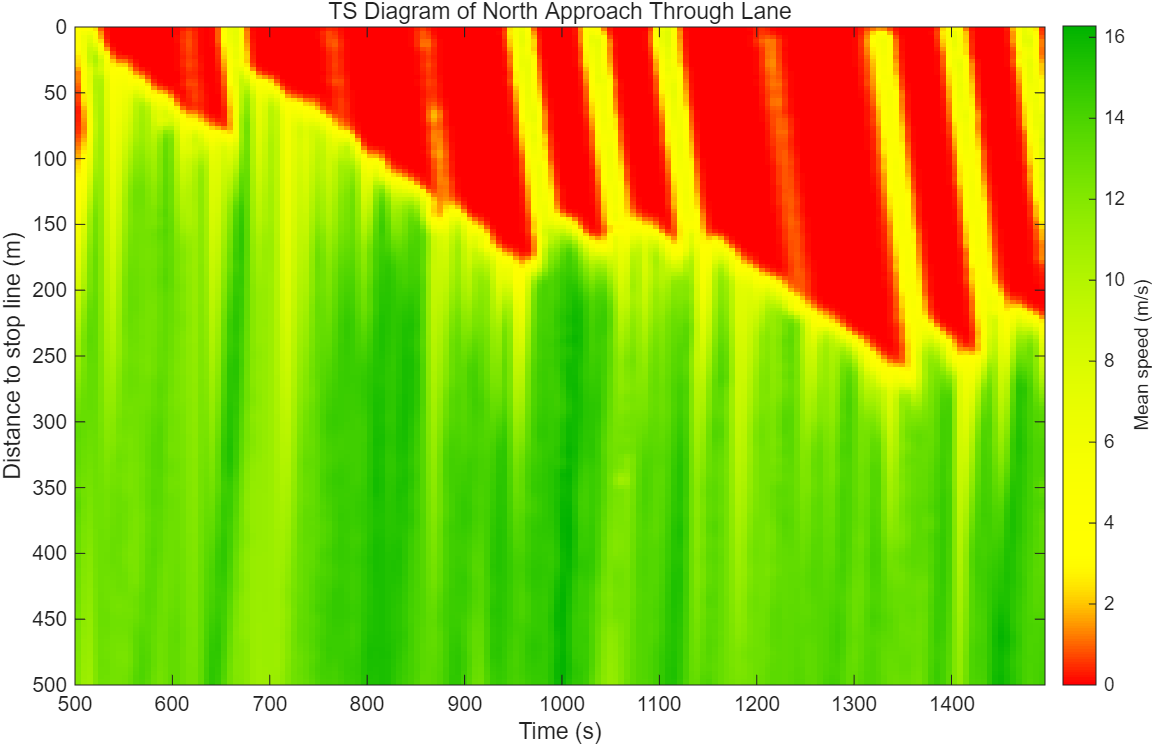}
        \caption{}
        \label{fig:ts_north_base}
    \end{subfigure}
    \hfill
    \begin{subfigure}{0.45\textwidth}
        \centering
        \includegraphics[width=\linewidth]{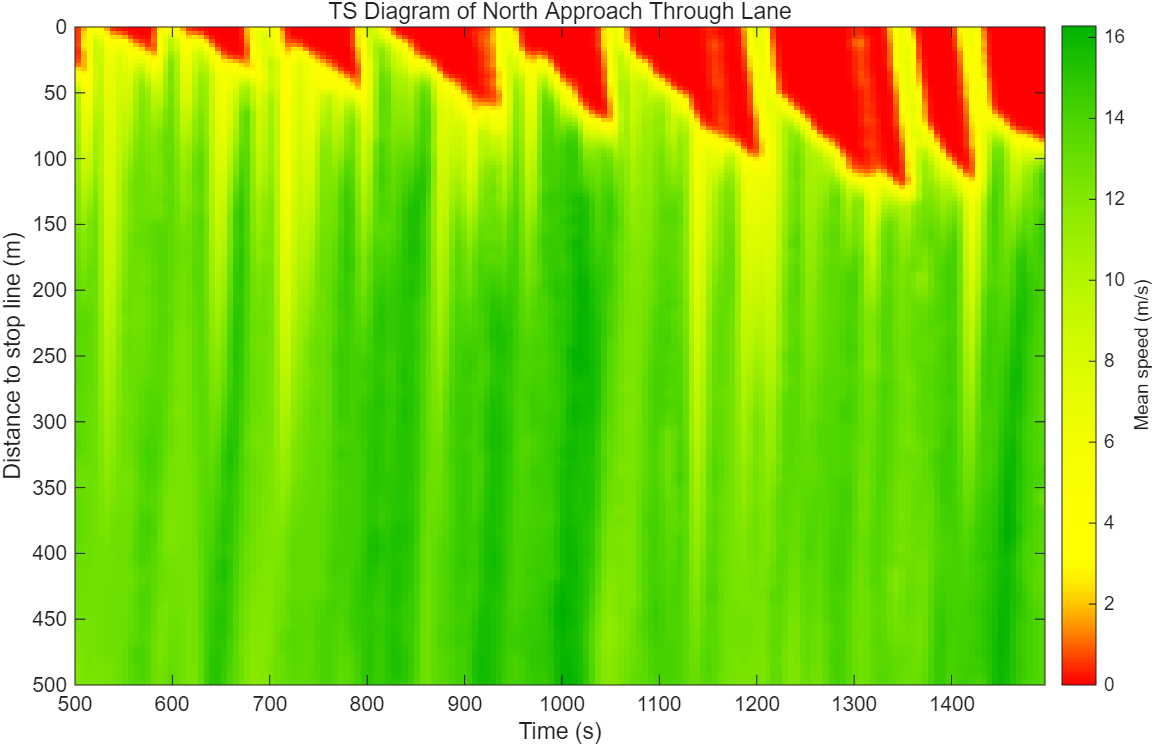}
        \caption{}
        \label{fig:ts_north_ours}
    \end{subfigure}

    \begin{subfigure}{0.45\textwidth}
        \centering
        \includegraphics[width=\linewidth]{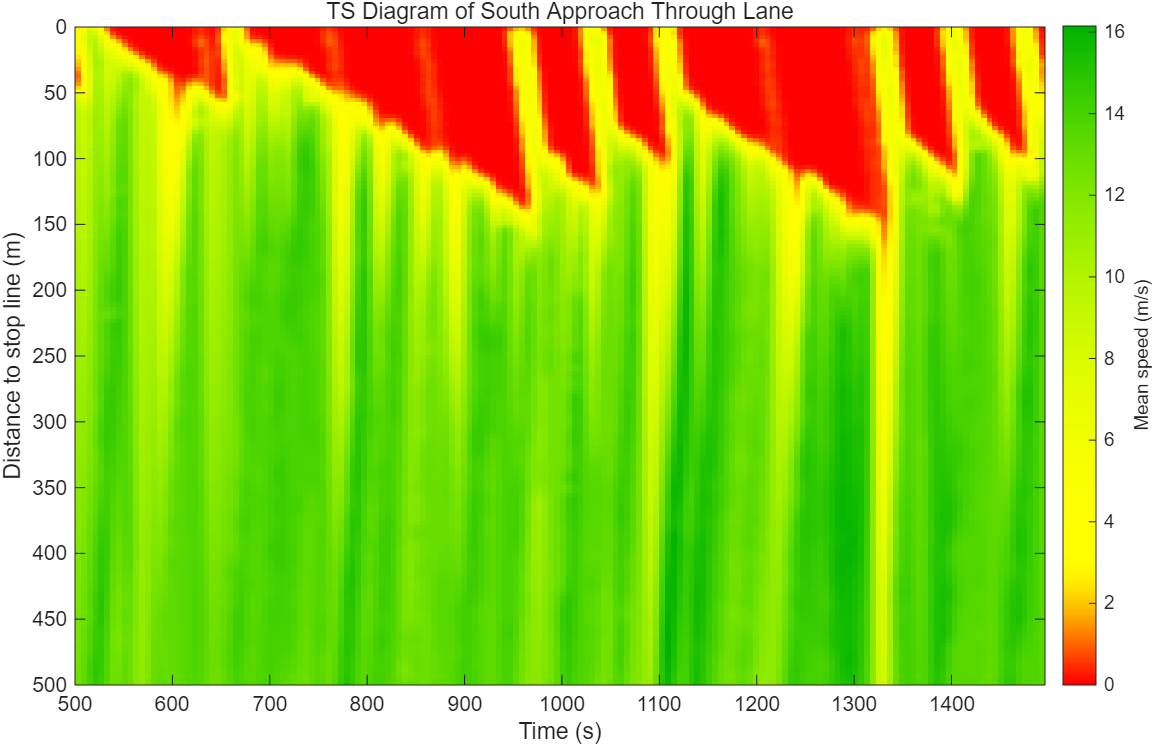}
        \caption{}
        \label{fig:ts_south_base}
    \end{subfigure}
    \hfill
    \begin{subfigure}{0.45\textwidth}
        \centering
        \includegraphics[width=\linewidth]{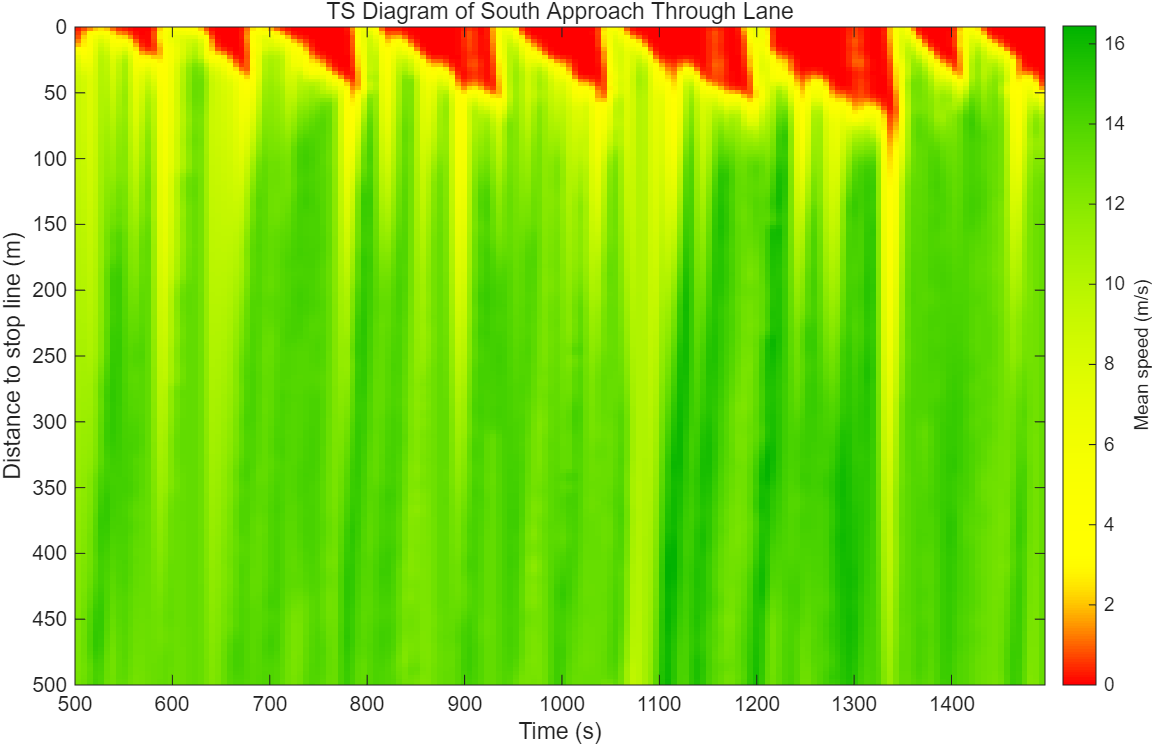}
        \caption{}
        \label{fig:ts_south_ours}
    \end{subfigure}
    
    \caption{Time–space diagrams of the north and south approach through lanes under the moderate-demand car accident scenario. Subfigures (a) and (c) correspond to the baseline controllers, while (b) and (d) illustrate the proposed TSC-LLM-TLRS. Compared with the baseline, the proposed framework substantially reduces the red congested regions.}
    \label{fig:ts}
    
\end{figure}

These improvements can also be visually confirmed in the time–space diagrams under the moderate-demand car accident scenario (see Figure \ref{fig:ts}). Specifically, we plot the trajectories on a north–south through lane, where the proposed TSC-LLM-TLRS substantially reduces congestion on the unaffected approaches. This indicates that the framework effectively enhances the efficiency of non-incident lanes while maintaining control over the disrupted traffic stream. Moreover, the overall system performance, in terms of both intersection-wide delay and queue length, still shows a consistent decrease (see Table \ref{tab:controller_comparison}), underscoring that the proposed framework not only mitigates localized disruptions but also enhances the resilience of the intersection to traffic incidents.

\noindent\textbf{Value of CoT and SRTLRS}: Beyond demonstration of LLM inference process and overall performance, it is important to highlight the value of the CoT reasoning module and SRTLRS. Although LLM-TSC without CoT performs comparably to LLM-TSC in the car accident scenario, a broader evaluation across both car accident and road maintenance cases reveals a consistent trend in which TSC-LLM outperforms TSC-LLM without CoT. This confirms that the CoT module is essential for ensuring logically consistent reasoning, as it helps the agent avoid overlooking affected lanes or assigning inappropriate fine-tuned parameters. Furthermore, comparing TSC-LLM with and without the TLRS shows that the retrieval mechanism provides an additional layer of robustness. By incorporating domain-specific knowledge from historical incident reports, the TLRS enables the agent to ground its reasoning in empirically validated practices, thereby producing strategies that are not only more reliable but also more aligned with fundamental traffic control principles. Together, these results demonstrate the complementary benefits of CoT reasoning and retrieval-augmented knowledge injection, highlighting their critical roles in enhancing the reliability of LLM-generated strategies.

\subsection{Performance comparison on incidents with no relevant references}\label{subsec:case-norel}
\begin{figure}
    \centering
    \includegraphics[width=1\linewidth]{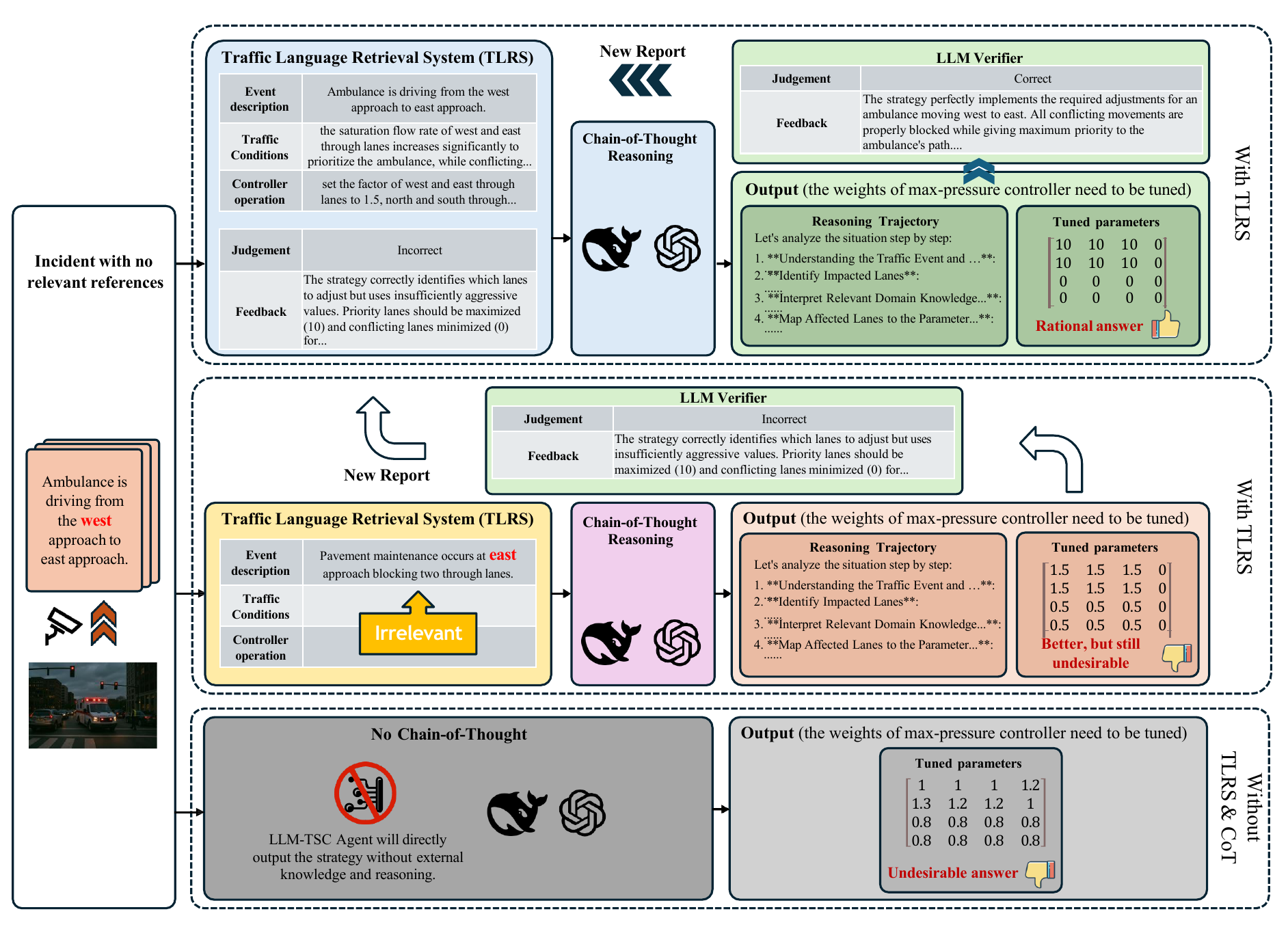}
    \caption{A case that demonstrates the inference process of the LLM Agent when handling incident with no relevant references.}
    \label{fig:case2}
\end{figure}

In this section, we focus on incidents with no relevant references in TLRS and follow the same evaluation approach as described in the previous section. Specifically, we examine the reasoning capability of the LLM Agent (see Figure \ref{fig:case2}), and similarly assess the resulting traffic performance by comparing four benchmarks using the corresponding evaluation metrics.

\noindent\textbf{Demonstration of LLM Inference Process:} Figure \ref{fig:case2} further illustrates a representative scenario in which an ambulance is traveling from the west approach to the east approach and requires prioritized passage. The MPC controller is employed as the lower-level controller. Given that this section focuses on incidents with no relevant references, we compare the inference processes of TSC-LLM without CoT and TSC-LLM-TLRS. For the latter, we present the agent’s outputs upon its first and second encounters with the incident.
Although the agent using TSC-LLM-TLRS fails to generate optimal fine-tuned parameters in the first encounter, it successfully produces an optimal strategy in the second encounter, aided by the verifier’s feedback. The agent correctly identifies the affected lanes and adjusts their parameters to appropriate values, which align with the intended control objectives. In contrast, for the benchmark TSC-LLM without CoT, the agent fails to generate a reasonable control strategy. It not only misidentifies the set of affected lanes but also provides insufficient parameter adjustments, falling short of achieving the control objectives required by this scenario.

\noindent\textbf{Performance Comparison:} Table \ref{tab:ambulance_elderly} presents the performance comparison among the benchmarks under two types of unforeseen traffic incidents, where no relevant references are available. In the elderly crossing scenario, since the focus lies solely on the crossing completion rate (CCR) of elderly pedestrians rather than vehicle performance, the experiments are conducted only under the oversaturated demand condition. The results in this section demonstrate the effectiveness of our framework in handling incidents with no relevant references, as evidenced by the significant improvements in both AD and CCR.

\begin{table}[htbp]
\centering
\footnotesize
\caption{Performance comparison under ambulance priority and elderly crossing scenarios. TSC-LLM-TLRS I denotes the performance of our method when it encounters a traffic incident for the first time, whereas TSC-LLM-TLRS II represents its performance during the second encounter with the same incident.}
\label{tab:ambulance_elderly}
\resizebox{\linewidth}{!}{
\begin{tabular}{@{}ccc >{\centering\arraybackslash}p{3.5cm} cc@{}}
\toprule
\textbf{Lower-level controller} & \textbf{Demand level} & \textbf{Methods} & \textbf{LLM} & \textbf{Ambulance Priority} & \textbf{Elderly Crossing} \\
& & & & AD (s) & CCR (\%) \\
\midrule

\multirow{14}{*}{\textbf{Max-pressure}}& \multirow{7}{*}{Oversaturated}& Baseline & - & 97.6 & 29.73\\
& & \multirow{2}{*}{TSC-LLM without CoT} & Deepseek-V3 & 65.1 & 4.05\\
& & & ChatGPT-4o & 71.5& 16.22\\
& & \multirow{2}{*}{TSC-LLM-TLRS I} & Deepseek-V3 & 5.2 & 84.46\\
& & & ChatGPT-4o & 71.0& 97.97\\
& & \multirow{2}{*}{TSC-LLM-TLRS II} & Deepseek-V3 & 0& 98.65\\
& & & ChatGPT-4o & 6.3& 98.65\\

\cmidrule(lr){2-6}
& \multirow{7}{*}{Moderate}& Baseline & - & 9.1& - \\
& & \multirow{2}{*}{TSC-LLM without CoT} & Deepseek-V3 & 12.5& - \\
& & & ChatGPT-4o & 11.1& \\
& & \multirow{2}{*}{TSC-LLM-TLRS I} & Deepseek-V3 & 3.2 & - \\
& & & ChatGPT-4o & 8.7& \\
& & \multirow{2}{*}{TSC-LLM-TLRS II} & Deepseek-V3 & 1.7& - \\
& & & ChatGPT-4o & 4.1& \\

\midrule

\multirow{14}{*}{\textbf{MPC}}& \multirow{7}{*}{Oversaturated}& Baseline & - & 50.4 & 18.92\\
& & \multirow{2}{*}{TSC-LLM without CoT} & Deepseek-V3 & 65.7 & 22.97\\
& & & ChatGPT-4o & 47.1& 27.70\\
& & \multirow{2}{*}{TSC-LLM-TLRS I} & Deepseek-V3 & 4.2 & 39.86\\
& & & ChatGPT-4o & 28.0& 86.49\\
& & \multirow{2}{*}{TSC-LLM-TLRS II} & Deepseek-V3 & 1.9 & 82.88\\
& & & ChatGPT-4o & 5.4& 96.62\\

\cmidrule(lr){2-6}
& \multirow{7}{*}{Moderate}& Baseline & - & 12.4& - \\
& & \multirow{2}{*}{TSC-LLM without CoT} & Deepseek-V3 & 7.8& - \\
& & & ChatGPT-4o & 12.8& \\
& & \multirow{2}{*}{TSC-LLM-TLRS I} & Deepseek-V3 & 3.2 & - \\
& & & ChatGPT-4o & 6.3& \\
& & \multirow{2}{*}{TSC-LLM-TLRS II} & Deepseek-V3 & 1.3& - \\
& & & ChatGPT-4o & 3.8& \\

\bottomrule
\end{tabular}
}
\end{table}
In the ambulance priority scenario, our method, \textbf{TSC-LLM-TLRS}, consistently outperforms all other methods across both demand levels and both lower-level controllers. Notably, under oversaturated demand with the max-pressure controller, the AD is reduced from 97.6 seconds to 0 seconds, indicating a complete elimination of intersection delay. Under all other conditions, our method also achieves substantial reductions in AD. Similarly, in the elderly crossing scenario, TSC-LLM-TLRS still achieves the best performance, significantly improving CCR compared to the baseline, reaching 80\% and 100\%, respectively.
By comparison, \textbf{TSC-LLM without CoT} exhibits unstable performance and does not show significant improvements over the baseline. In some cases, such as under the MPC controller, it even leads to a slight increase in ambulance delay. This further highlights the necessity of incorporating the CoT module for enhancing the LLM Agent’s reasoning capabilities.

\noindent\textbf{Value of Self-Refinement:} Table \ref{tab:ambulance_elderly} shows a significant decrease in average delay and an increase in CCR when moving from TSC-LLM-TLRS I to TSC-LLM-TLRS II. 
For example, the CCR of elderly crossing increases from 84.46\% to 98.65\% with 16.8\% improvement for the max-pressure controller augmented by Deepseek-V3.
This indicates that while TSC-LLM-TLRS may not always produce an optimal strategy during its first encounter with an incident due to the absence of relevant references, the framework rapidly improves in subsequent iterations.
Specifically, the LLM-based verifier enables the agent to accurately evaluate its own performance, and the resulting feedback is incorporated as new experience into the TLRS. When the incident recurs, the agent can leverage accumulated experience to adopt appropriate measures and avoid suboptimal actions. This learning process closely resembles how an experienced traffic police officer handles unforeseen incidents by continuously learning from past cases in real-world operations.
Overall, these results demonstrate the effectiveness of the self-refined TLRS module, which integrates retrieved historical reports with verifier feedback. Such self-refinement ensures that even in the absence of relevant prior cases in TLRS, the framework can still generate reliable outputs with strong generalizability.

\section{Conclusion} \label{sec-con}
In this paper, we propose a novel LLM-augmented TSC framework that enhances the reliability of conventional adaptive TSC methods in response to unforeseen traffic incidents. We design an LLM agent that can perform as virtual traffic police officers at the upper level to fine-tune the controller parameters at the lower level. This framework provides valuable LLM-based guidance as an add-on to existing controllers, enhancing their ability to generalize to traffic incidents without incurring high deployment costs. To enhance the reliability of the LLM agent’s outputs, we devise two mechanisms to enhance its domain knowledge. First, we develop a TLRS that provides relevant expert knowledge to guide the LLM’s inference process. Second, we introduce a self-refinement mechanism in which an LLM-based verifier continuously evaluates the agent’s outputs and updates the TLRS over time. These mechanisms enable the LLM agent to reason in a manner similar to an experienced traffic police officer.
Simulation results under a wide range of traffic incident scenarios demonstrate that the proposed method significantly outperforms both conventional adaptive controllers and LLM-based controllers without TLRS support. These results suggest that effective strategies exist to ensure reliable and trustworthy LLM outputs for safety-critical traffic environments.

This work opens several promising future directions.
First, the TLRS can be extend to graph-based representations \citep{edge2024local} to better capture the traffic languages in historical incidents reports and further improve the reasoning process of the agent.
Second, we would like to develop a safety filter \citep{kumar2023certifying} to provide guarantees for traffic control performance.
Third, the simulations can be extend to large-scale urban networks with more diverse traffic incident types to support more comprehensive real-world deployment. 
\printcredits

\section*{Acknowledgment}
This research was supported by the Singapore Ministry of Education (MOE) under its Academic Research Fund Tier 1 (A-8001183-00-00).

\section*{Authorship Contribution Statement}
\textbf{Shiqi Wei}: Methodology, Validation, Writing - original draft. 

\textbf{Kaidi Yang}: Conceptualization, Methodology, Writing - original draft, revised draft. 
\bibliographystyle{cas-model2-names}

\bibliography{cas-refs}

\end{document}